\documentclass[12pt]{article}
\usepackage[english]{babel}
\usepackage{amsmath,amssymb}
\usepackage{cite,./mcite}
\usepackage{array}
\usepackage{xspace}
\usepackage{endnotes}
\usepackage{a4,epsfig}
\usepackage{rotate}
\usepackage{./tbookdef}
\begin{document}
%
%
 
\begin{titlepage}
 
\title{\Large{\centerline{\bf Heavy Quark Production Measurements
at THERA~\footnote{To appear in ``The THERA Book", DESY-LC-REV-2001-062 (2001). 
}
}}}
\date{15th May 2001}

\maketitle

\vspace*{1cm}
\begin{center}
 L. Gladilin\\
 {\it Weizmann Institute, Department of Particle Physics, \\
 Rehovot, Israel\\
 on leave from Moscow State University\\
 gladilin@mail.desy.de }\\~\\
 I. Redondo\\
 {\it Universidad Aut\'onoma de Madrid,\\
 Departamento de F\'{\i}sica Te\'orica, Madrid, Spain\\
 redondo@mail.desy.de}
\end{center}
\vspace*{1cm}

\centerline{\bf Abstract}
\vspace*{1cm}

The total cross sections of charm and beauty production
at a future high--energy $ep$ collider, THERA, are expected
to increase by factors of three and five, respectively,
as compared to HERA.
Heavy quarks can be measured at THERA in wide ranges of
transverse momenta and $Q^2$ values,
thereby providing a solid basis for testing the perturbative
QCD calculations.
The charm and beauty contributions to the proton structure can be probed
at THERA at $\sim1$ order
of magnitude smaller Bjorken $x$ values with respect to those at HERA.
Charm production in the process of photon--gluon fusion at THERA
can serve for the determination of the gluon structure of the proton
in the as yet unexplored kinematic range $10^{-5}<x_g<10^{-4}$.
The cross section of charm production in charged current at THERA
is $\sim 1$ order of magnitude larger than that at HERA.
 
\setcounter{page}{1}
\thispagestyle{empty} 

\end{titlepage}
 

\section{Introduction}

Heavy quarks are produced copiously at HERA which provides
collisions between electrons or positrons with
energy $E_e=27.5\,\gev$ and protons with energy
$E_p=920\,\gev$~\footnote{
The proton energy was $820\,\gev$ from 1992 to 1997.}.
The total charm and beauty cross sections at HERA
are of the order of $1\,\mu\text{b}$ and
$10\,\text{nb}$, respectively~\cite{pl:b308:137}.
In photoproduction processes at HERA, a quasi-real photon
with virtuality $Q^2\sim0$ is emitted by the incoming electron
and interacts with the proton. At leading order (LO) in QCD
two types of processes are responsible for the production
of heavy quarks: the direct photon processes, where the photon
participates as a point-like particle, and the resolved photon
processes, where the photon acts as a source of partons. The dominant
direct photon process is photon--gluon fusion (PGF) where
the photon fuses with a gluon from the incoming proton.
In resolved photon processes, a parton from the photon scatters
off a parton from the proton.
Charm and beauty quarks present in the parton distributions
of the photon, as well as of the proton, lead to processes like
$cg \rightarrow cg$ and $bg \rightarrow bg$,
which are called heavy flavour excitation processes.
In next-to-leading order (NLO) QCD only the sum of direct
and resolved processes is unambiguously defined.
The resolved photon processes are suppressed
in deep inelastic scattering (DIS) at HERA because the virtuality of
the exchanged photon is typically selected to be $Q^2>1\,\gev^2$.

Charm production at HERA has been studied by the
H1 and ZEUS collaborations in both the photoproduction and DIS
regimes~\cite{np:b545:21,epj:c6:67,epj:c12:35,pl:b481:213}.
A description of the charm photoproduction cross sections is rather
problematic for present perturbative QCD (pQCD) calculations.
The fixed-order NLO calculations~\cite{pl:b348:633,*np:b454:3} are
generally below the measured
cross sections, in particular in the forward (proton)
direction~\cite{epj:c6:67,pl:b481:213}. The fixed-order approach assumes
that gluons and light quarks (u,d,s) are the only active partons
in the structure functions of the proton and the photon. In this approach
there is no explicit heavy flavour excitation component and
heavy quarks are produced only dynamically in hard pQCD processes.
The resummed NLO
calculations~\cite{zfp:c76:689,pr:d58:014014,pr:d55:2736,*pr:d55:7134}
treat charm as an additional active parton in the structure functions.
These calculations are
valid only if the heavy quark transverse momentum is much larger
than $M_{c,b}$, where $M_{c,b}$ is the charm or beauty quark mass.
The resummed NLO predictions of~\cite{zfp:c76:689,pr:d58:014014}
are rather close to the measured cross sections~\cite{epj:c6:67}.

The fixed-order or three flavour
Fixed Flavour Number Scheme (FFNS) calculations
for charm production in DIS~\cite{pr:d57:2806}
agree with the cross sections measured at HERA~\cite{np:b545:21,epj:c12:35}.
These calculations are expected to be less reliable when
$Q^2/M_c^2\gg 1$.
In this range the $\ln(Q^2/M_c^2)$ terms should be resummed
and absorbed into the charm distribution function in
the proton~\cite{pr:d50:3102,jhep:0010:031,pr:d61:096004}.

The first measured beauty photoproduction cross sections at
HERA~\cite{pl:b467:156,epj:c18:625} lie above the fixed-order
NLO QCD
predictions~\cite{np:b412:225,*np:b373:295,pl:b348:633,*np:b454:3}.
The preliminary value of the beauty production cross section
in the DIS regime at HERA,
reported recently by the H1 collaboration~\cite{h1-bdis},
exceeds the FFNS prediction~\cite{pr:d57:2806}.

Future THERA collider will utilize proton beam from HERA and electron
beam prepared with one or both arms of the TESLA $e^+e^-$ linear
collider. Using only one arm of the electron accelerator,
electron energies of $250\,\gev$ and $400\,\gev$ can be reached
in the first and second stages of TESLA, respectively.
The electron energy can be as large as $800\,\gev$ employing
both TESLA arms.
An increase of the centre-of-mass energy of $ep$ collisions from about
$300\,\gev$ at HERA to $\sim1\,\tev$ at THERA will result in quite
significant increase of the
total heavy quark production cross sections.
Tables~\ref{tab:hqpm:direct} and ~\ref{tab:hqpm:resolved}
show LO estimations of the cross sections in
direct and resolved photon processes, respectively.
The estimations were obtained with the LO Monte Carlo program
HERWIG~\cite{cpc:67:465}.

\begin{table}[hbt]
  \begin{center}
    \begin{tabular}{|c|c|c|c|} \hline
      Collider & charm [$\mu\text{b}$] &  beauty [$\text{nb}$] &  top [$\text{fb}$] \\
      \hline
      \hline
      HERA, $E_e=27.5\,\gev$ & 0.6 & 4.3 & -- \\
      \hline
      THERA, $E_e=250\,\gev$ & 1.6 & 17 & 6.9 \\
      \hline
      THERA, $E_e=400\,\gev$ & 1.9 & 22 & 26 \\
      \hline
      THERA, $E_e=800\,\gev$ & 2.4 & 32 & $1.2\cdot10^2$ \\
      \hline
    \end{tabular}
  \end{center}
  \vskip-0.5cm
  \caption{
  Cross sections of charm, beauty and top production in direct photon
  processes at HERA and THERA estimated with the LO Monte Carlo program HERWIG.}
  \label{tab:hqpm:direct}
\end{table}
\vskip-0.5cm
\begin{table}[hbt]
  \begin{center}
    \begin{tabular}{|c|c|c|c|} \hline
      Collider & charm [$\mu\text{b}$] &  beauty [$\text{nb}$] &  top [$\text{fb}$] \\
      \hline
      \hline
      HERA, $E_e=27.5\,\gev$ & 0.3 & 1.4 & -- \\
      \hline
      THERA, $E_e=250\,\gev$ & 1.2 & 14 & 0.2 \\
      \hline
      THERA, $E_e=400\,\gev$ & 1.6 & 21 & 0.6 \\
      \hline
      THERA, $E_e=800\,\gev$ & 2.5 & 38 & 1.9 \\
      \hline
    \end{tabular}
  \end{center}
  \vskip-0.5cm
  \caption{
  Cross sections of charm, beauty and top production in resolved photon
  processes at HERA and THERA estimated with the LO Monte Carlo program HERWIG.}
  \label{tab:hqpm:resolved}
\end{table}

The total charm and beauty production cross sections at THERA
with $E_e=250\,\gev$ are larger than those at HERA
by factors $\sim3$ and $\sim5$, respectively.
They grow further with the increase of the electron beam energy.
The growth is larger for heavy quark production
in the resolved photon processes.
The total cross sections of top quark production
at THERA with $E_e=250\,\gev$ and $E_e=800\,\gev$
are of the order of $10\,\text{fb}$ and $100\,\text{fb}$, respectively.

\section{Heavy quark photoproduction and proton gluon structure}

Charm and beauty production in $ep$ collisions is dominated by
photoproduction with $Q^2\sim0$. In this regime both direct and resolved
photon contributions are sizable. Prospects for investigations
of heavy quark production in resolved photon processes at THERA
are discussed elsewhere in this book~\cite{thera-jankowski}.
In this section we will discuss charm and beauty photoproduction
in direct photon processes and its sensitivity to the gluon structure
of the proton.

\begin{figure}[hbt]
  \begin{center}
    \epsfig{figure=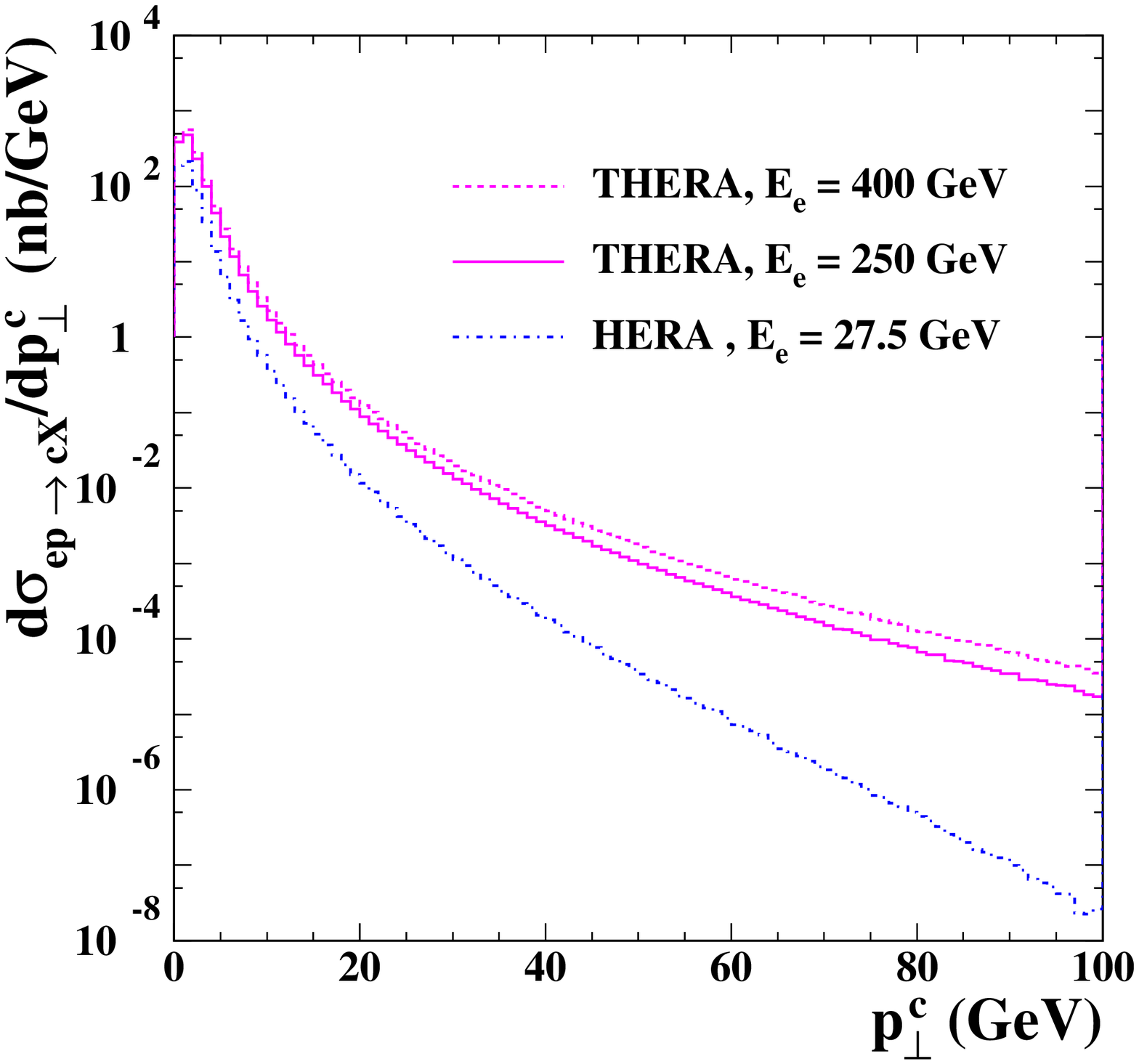,width=7.5cm}
    \epsfig{figure=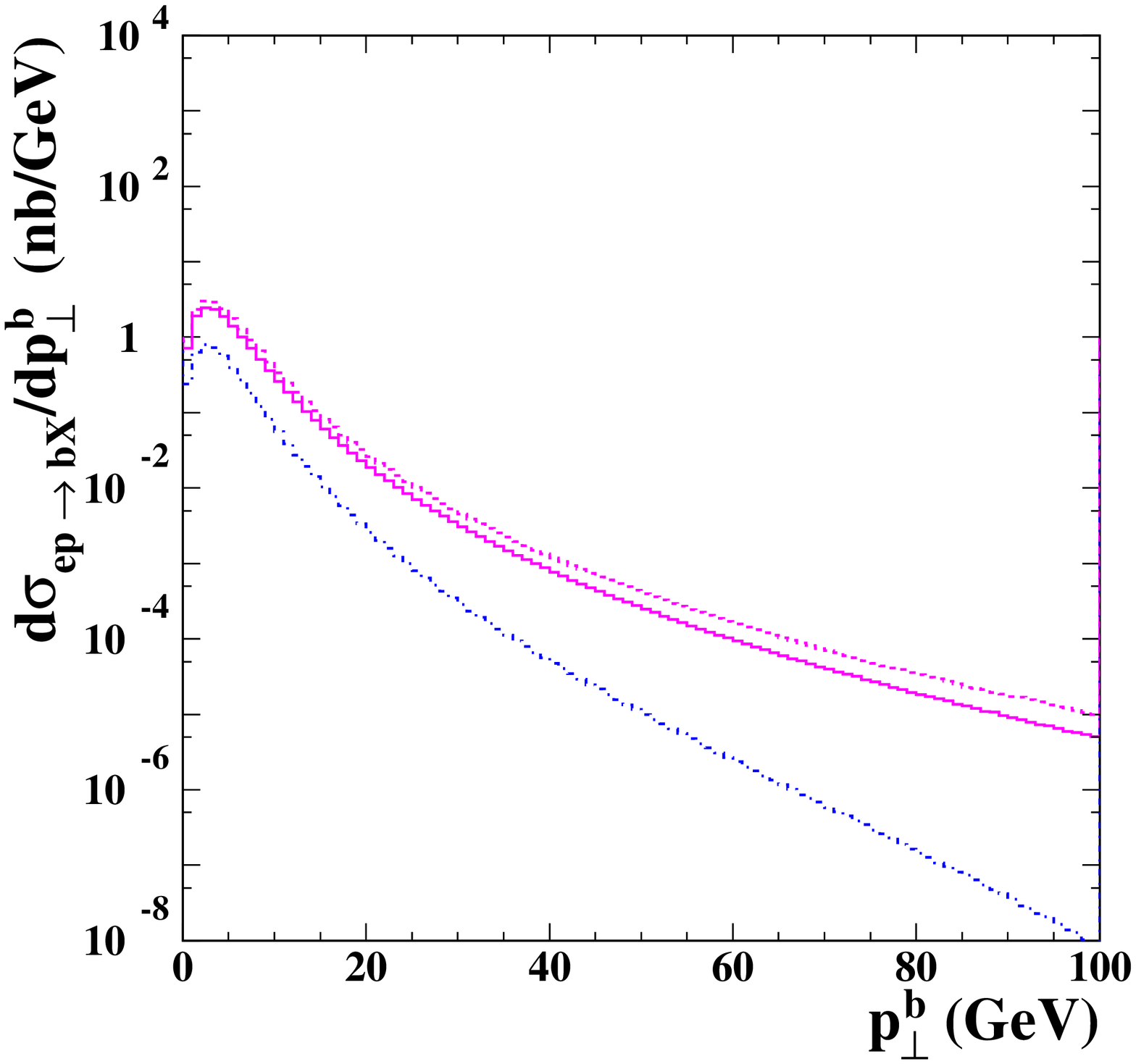,width=7.5cm}
  \end{center}
  \vskip-2.5cm
  \centerline{\large\kern1.7cm{\bf a)\ }$\gamma g\to c\widebar c$
              \hfill{\bf b)\ }$\gamma  g\to b\widebar b$\kern3.6cm}
  \vskip1.0cm
\begin{center}
\epsfig{figure=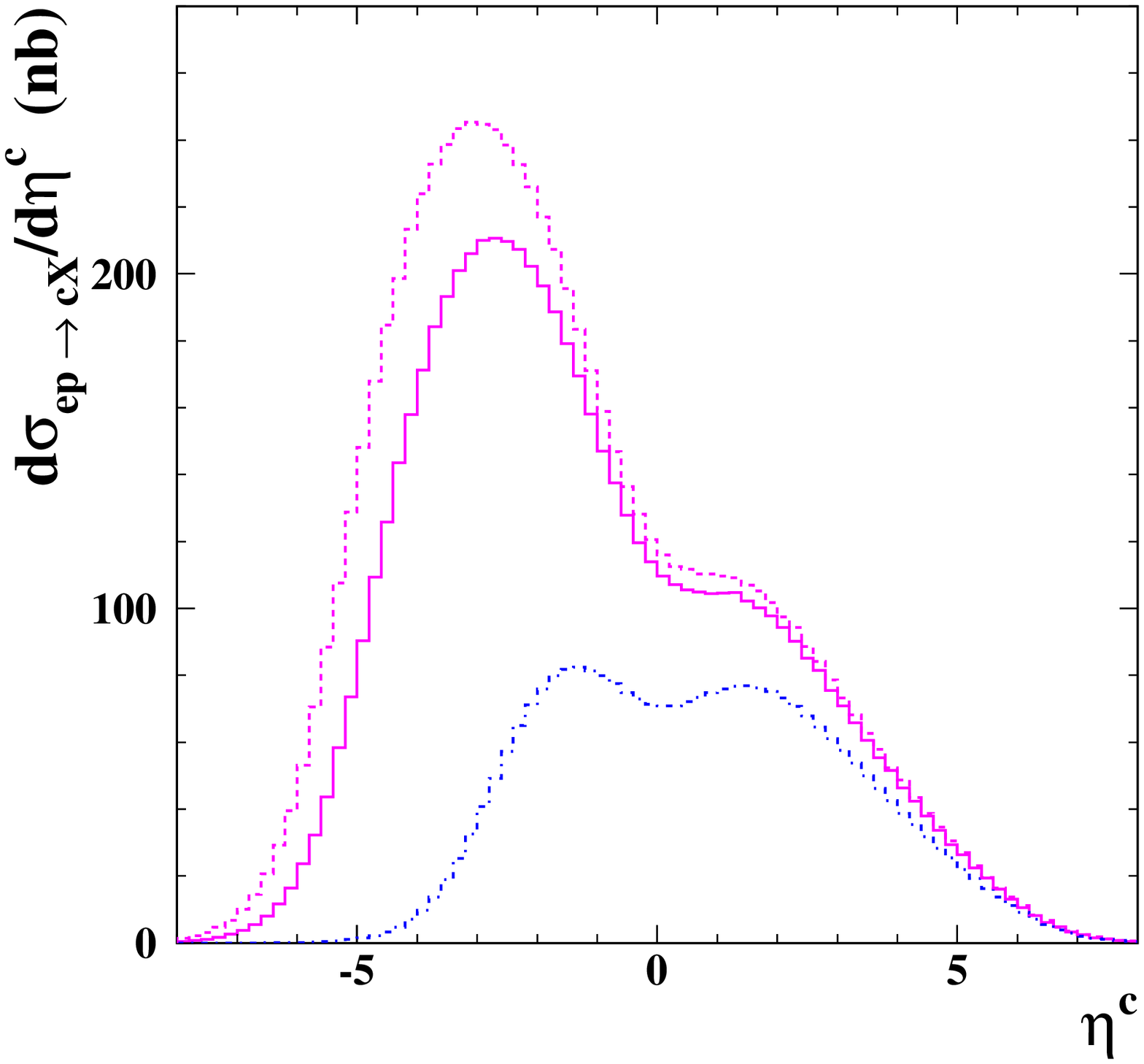,width=7.5cm}
\epsfig{figure=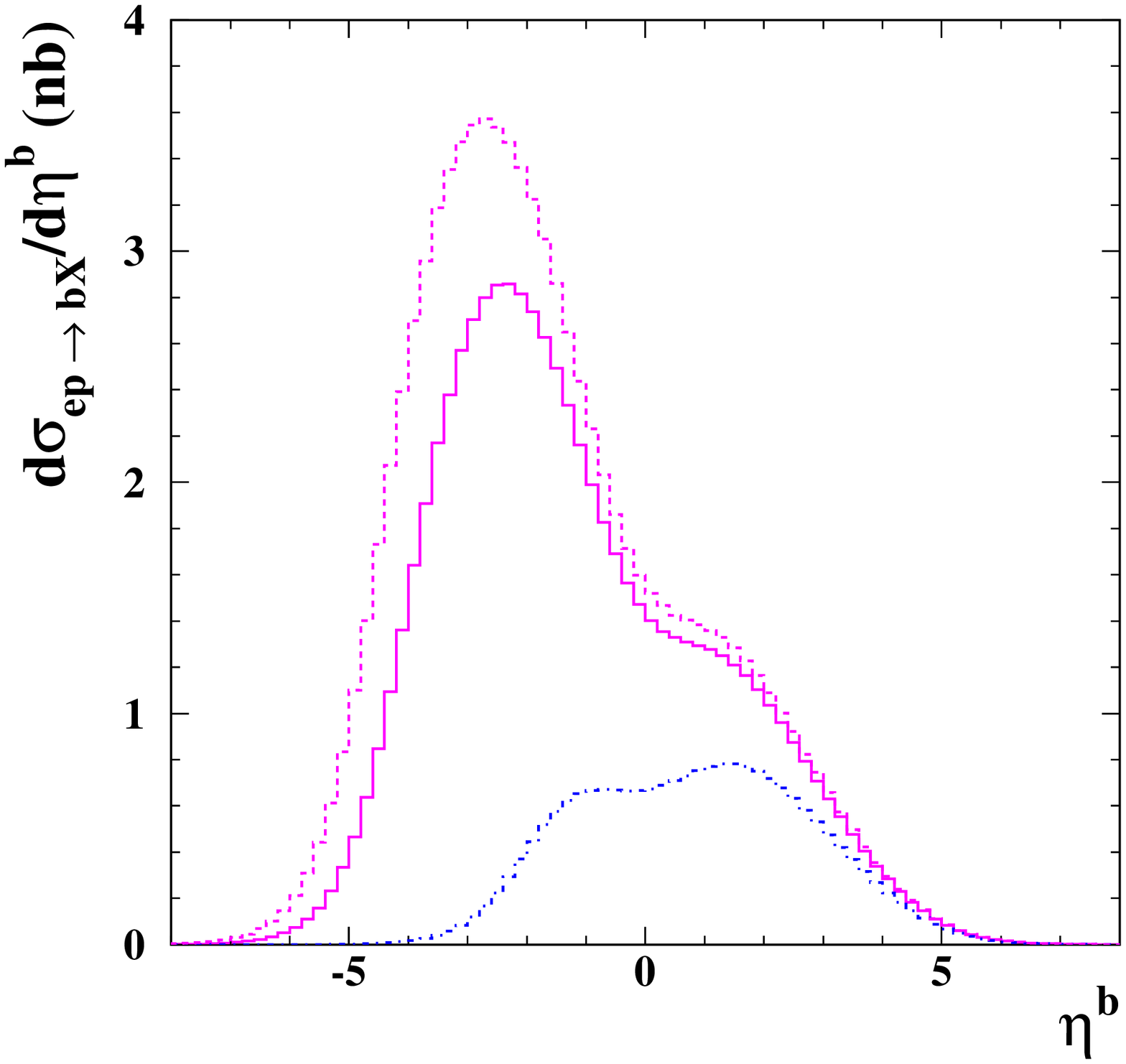,width=7.5cm}
\end{center}
  \vskip-6.5cm
  \centerline{\large\kern4.2cm{\bf c)\ }$\gamma g\to c\widebar c$
              \hfill{\bf d)\ }$\gamma  g\to b\widebar b$\kern1.1cm}
  \vskip5.0cm
  \caption{
    The contribution of photon--gluon fusion to the differential cross
    sections $d\sigma/dp_\perp$ and $d\sigma/d\eta$ for
    charm ((a) and (c)) and
    beauty ((b) and (d)) production
    calculated in NLO
    QCD for $Q^2<1\,\gev^2$. The solid and dashed magenta curves show the
    predictions for THERA operation with an electron energy of $250\,\gev$ and
    $400\,\gev$, respectively, and $E_p=920\,\gev$. The predictions
    for the HERA
    case are indicated by the dash-dotted blue curves.}
  \label{fig:hqpm:bgf}
  \vskip-5.mm
\end{figure}

Fig.~\ref{fig:hqpm:bgf} shows the contributions of photon--gluon fusion to
the differential cross sections $d\sigma/ dp^{c,b}_\perp$
and $d\sigma/ d\eta^{c,b}$
($p^{c,b}_\perp$ and $\eta^{c,b}$ denoting the quark transverse momentum
and pseudorapidity~\footnote{
The pseudorapidity $\eta$ is defined as $-\ln(\tan\frac{\theta}{2})$, where
the polar angle $\theta$ is taken with respect to the proton
beam direction.})
at HERA and THERA, calculated
within NLO QCD~\cite{np:b412:225,*np:b373:295} for $Q^2<1\,\gev^2$.
The difference between the heavy quark production cross sections at THERA
and HERA increases with increasing $p^{c,b}_\perp$, thereby creating
the opportunity to measure charm and beauty quarks at THERA in a wider
transverse momentum range. Such measurements will provide a solid basis
for testing the fixed-order, resummed, and
$k_t$-factorization~\cite{thera-baranov} pQCD calculations.
One has to note that the heavy quark pseudorapidity
distributions are shifted to backward (electron)
direction at THERA with respect to those at HERA.
\begin{figure}[hbt]
  \begin{center}
    \epsfig{figure=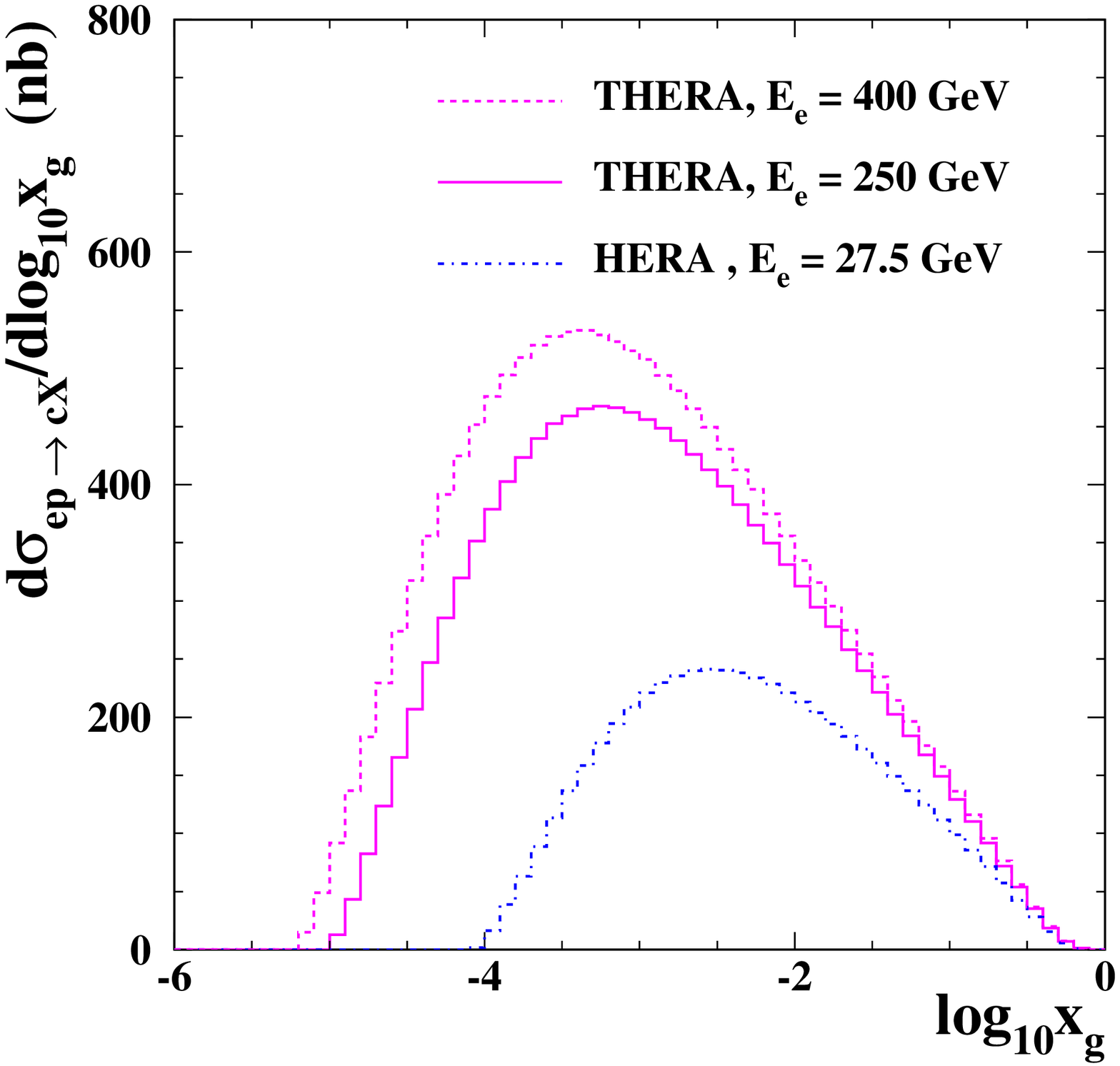,width=7.5cm}
    \epsfig{figure=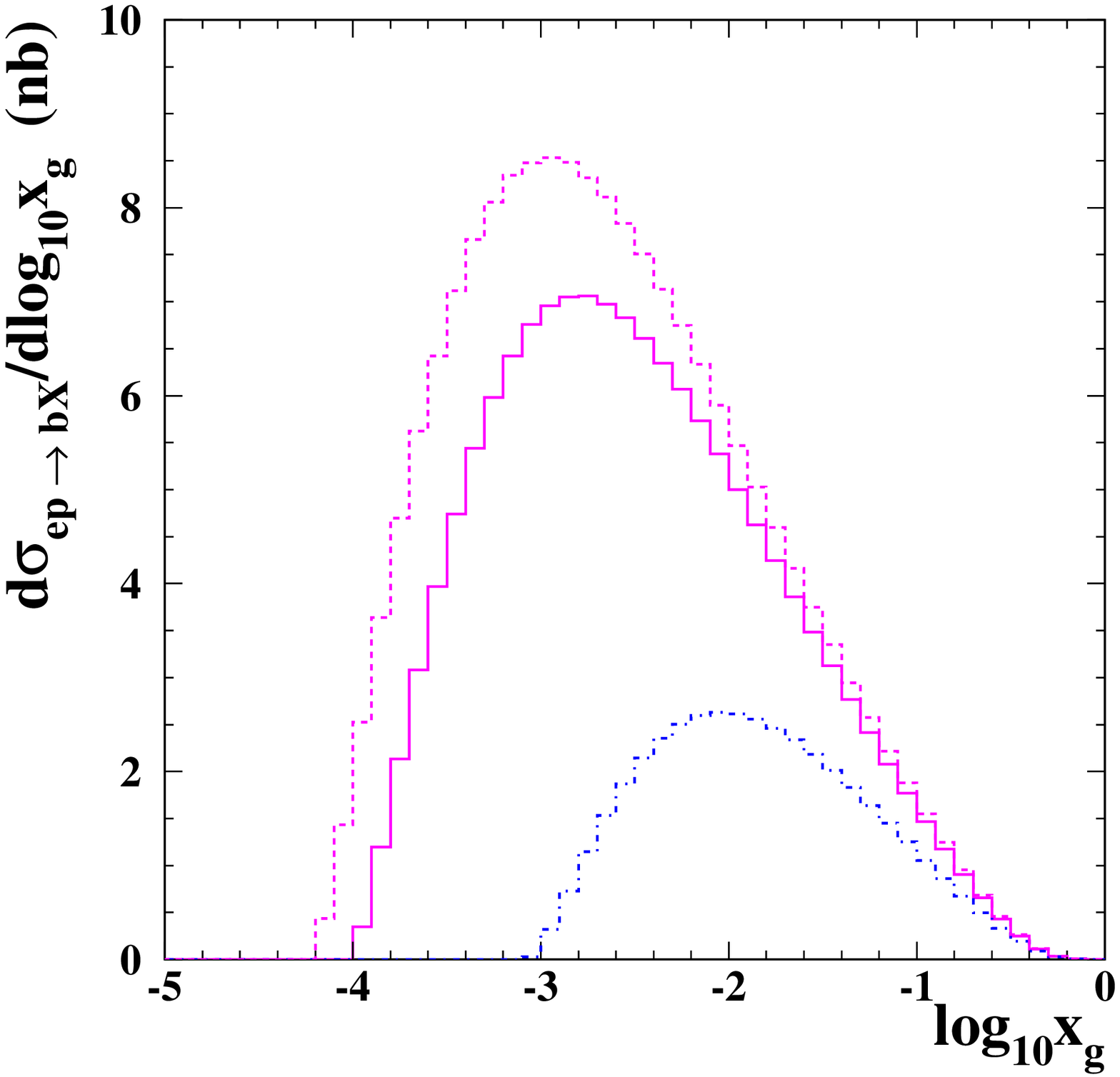,width=7.5cm}
  \end{center}
  \vskip-6.5cm
  \centerline{\large\bf\kern1.7cm a)\kern7.5cm b)\hfill}
  \vskip5.5cm
  \begin{center}
    \epsfig{figure=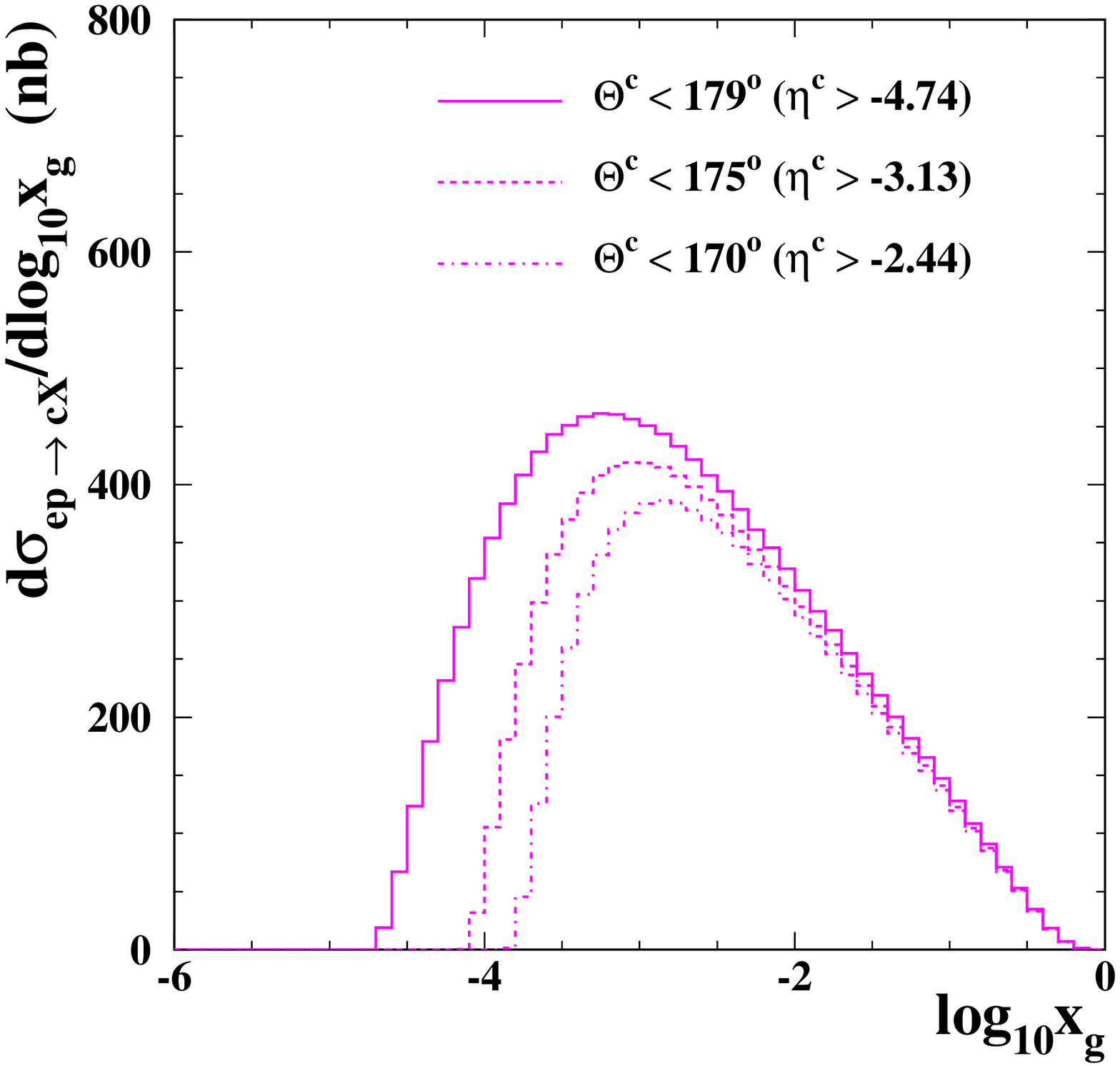,width=7.5cm}
    \epsfig{figure=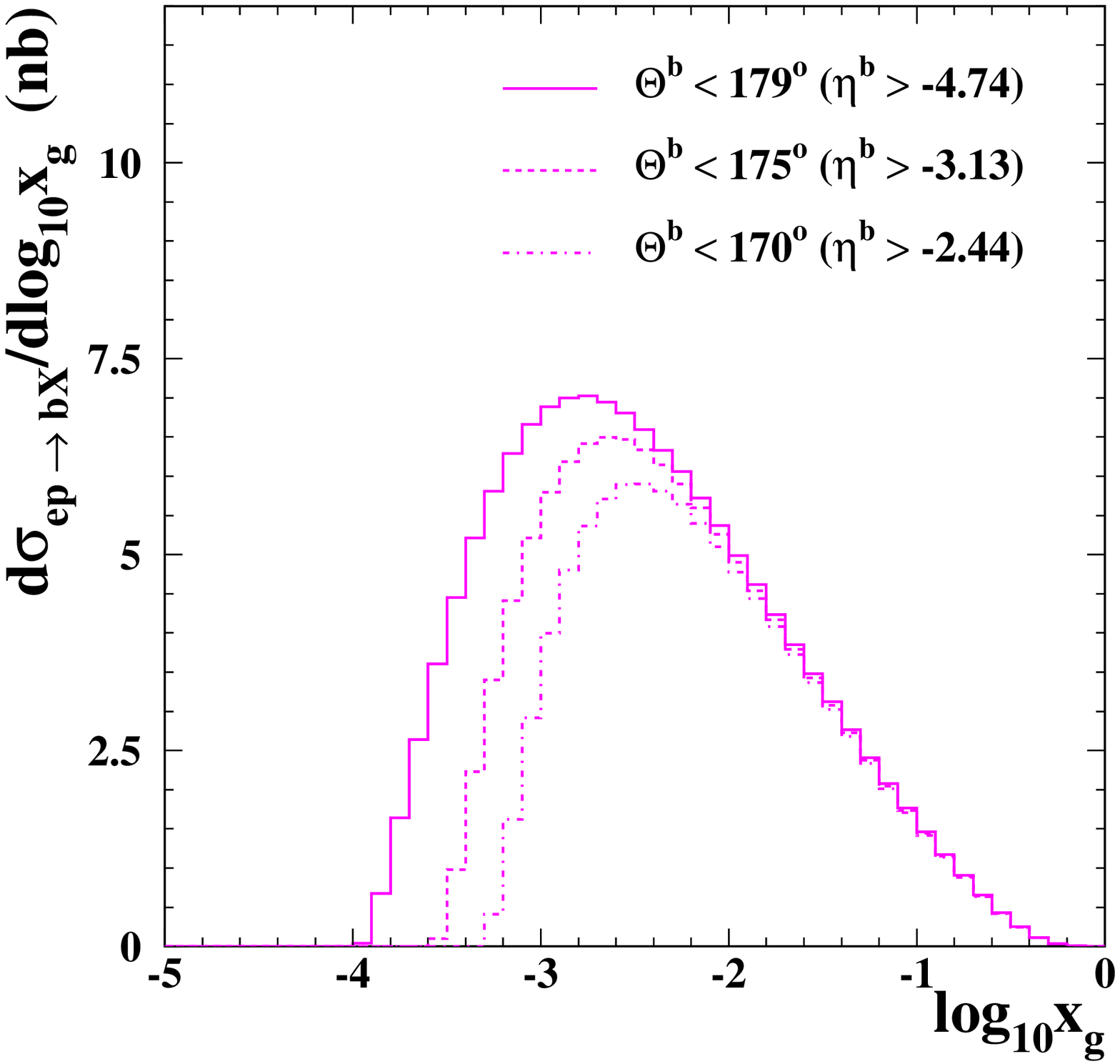,width=7.5cm}
  \end{center}
  \vskip-6.5cm
  \centerline{\large\bf\kern1.7cm c)\kern7.5cm d)\hfill}
  \vskip5.3cm
  \caption{The differential cross sections
  $d\sigma/d\log_{10}x_g$ for charm ((a) and (c)) and
  beauty ((b) and (d)) produced in the process of photon--gluon fusion.
  The cross sections were calculated within
  NLO QCD for $Q^2<1\,\gev^2$.
  In (a) and (b), the solid and dashed magenta curves show the
  predictions for THERA operation with an electron energy of $250\,\gev$ and
  $400\,\gev$, respectively, and $E_p=920\,\gev$. The predictions
  for the HERA case are indicated by the dash-dotted blue curves.
  In (c) and (d), the predictions for THERA with $E_e=250\,\gev$
  are shown with additional cuts $\theta^{c,b}<179^\circ$ (solid curves),
  $\theta^{c,b}<175^\circ$ (dashed curves) and
  $\theta^{c,b}<170^\circ$ (dash-dotted curves).}
  \label{fig:hqpm:xglu_bgf}
  \vskip-5.mm
\end{figure}
To measure charm and beauty hadronization products
a detector for THERA should be equiped with
special tracking and muon identification devices
in the backward direction.

Measurements of the heavy quarks produced in the process of
photon--gluon fusion can be used for the direct reconstruction
of the gluon structure of the proton~\cite{np:b545:21}.
Fig.~\ref{fig:hqpm:xglu_bgf} shows the differential cross sections
$d\sigma/d\log_{10}x_g$
($x_g$ denoting the gluon fractional momentum in the proton)
for charm and beauty produced in PGF
at HERA and THERA.
The cross sections were calculated within
NLO QCD~\cite{np:b412:225,*np:b373:295} for $Q^2<1\,\gev^2$.
The increase of the electron beam energy
will provide an opportunity to probe at THERA one order of magnitude
smaller $x_g$ values with respect to those at HERA.
The kinematic limits of the $x_g$ measurements at THERA are
$10^{-5}$ and $10^{-4}$ for charm and beauty production, respectively.
However, to be sensitive to the $x_g$ values around
the kinematic limits one will need to tag heavy quarks in
the very backward direction at THERA.
Plots (c) and (d) in Fig.~\ref{fig:hqpm:xglu_bgf} show
the predictions for THERA with $E_e=250\,\gev$
imposing additional cuts $\theta^{c,b}<179^\circ$,
$\theta^{c,b}<175^\circ$ and $\theta^{c,b}<170^\circ$.
Only charm quarks with $\theta^c>175^\circ$ demonstrate
sensitivity to the as yet unexplored range
$10^{-5} < x_g < 10^{-4}$.

\clearpage

\section{Heavy quark production in neutral current DIS}

Electron--proton scattering at $\sqrt{s} \sim1\,\tev$ will open new regions 
for heavy quark production in DIS never measured before. 
One of the legacies of HERA is the experimental confirmation that
charm electroproduction 
in the HERA regime is dominated by PGF~\cite{np:b545:21,epj:c12:35}.
It is, therefore, a direct probe of the gluon in the proton. 

A NLO QCD Monte Carlo program HVQDIS~\cite{pr:d57:2806} 
was used  to study  the sensitivity for  charm and bottom 
production  at THERA
with respect to that at HERA
focusing in the increased reach at small Bjorken $x$ values.
The HVQDIS program  produces fully differential distributions
in the heavy quark momenta.
It implements three flavour FFNS matrix elements
to order $\alpha_s^2$ calculated previously
in~\cite{np:b392:162,*pl:b353:535},
and gives a fair description of charm electroproduction
at HERA~\cite{epj:c12:35}.

\begin{figure}[hbt]
  \begin{center}
    \epsfig{figure=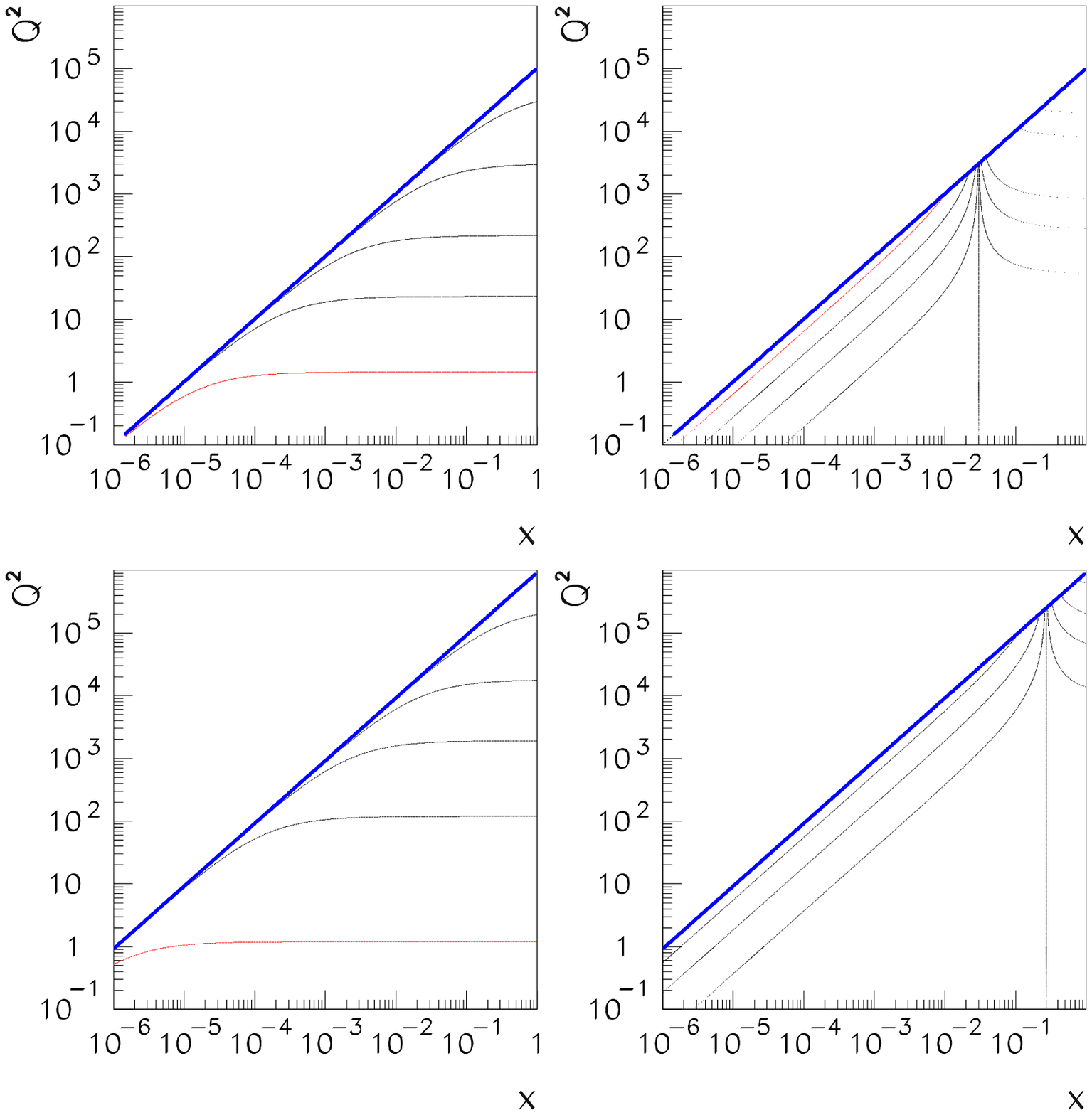,width=16cm}
  \end{center}
  \caption{
  The scattered electron angle (left) and energy (right) constant lines
  in the $Q^2$ and $x$ plane. Top (bottom)  plots are
  for HERA  (THERA) kinematics.  Thick (blue) diagonal lines are
  the kinematic limits.
  The values of constant angle  and energy  lines plotted for HERA(THERA)
  kinematics  are 177.5, 170, 150, 90, 30
  (179.75, 177.5, 170, 150, 90) degrees and 10, 20,
  25, 27, 27.5, 28, 30, 35, 100, 200  (10, 100, 200, 240, 250, 260,
  300, 400, 500) GeV. 
  The 10 GeV line is indistinguishable from
  the kinematic limit in the case of THERA.}
  \label{fig:hqpm:kplane}
  \vskip-5.mm
\end{figure}

It was  assumed that scattered electrons will be measurable down to
very low angles
allowing sizable acceptance for DIS events with $Q^2>1\,\gev^2$.  
Fig.~\ref{fig:hqpm:kplane} shows
the scattered electron angle  and energy  constant lines
for HERA and THERA.
The angle line corresponding to
the current ZEUS detector limit (177.5 degrees) excludes most of
the region bellow $Q^2=100\,\gev^2$ at THERA.
Thus, to measure scattered electrons in the low $Q^2$ DIS range
one will need a special low angle detector at THERA.

Minimum cuts on the
$p_\perp$ and $\eta$ of
the  heavy quarks were imposed
to take into account the detector acceptance for the heavy quark
tagging.
The choices made here are an educated guess, which can be considered
optimistic.

The following kinematic range was selected:
\begin{itemize}
\item    $Q^2 > 1\,\gev^2$ 
\item    $p^{c,b}_\perp >3\,\gev$ 
\item    $\mid{\eta^{c,b}} \mid<5$ 
\end{itemize}

HVQDIS was run with the following settings unless otherwise stated: 
\begin{itemize}
\item        $E_{e}$ = $250\,\gev$ and $E_{p}$ = $1\,\tev$ for THERA
\item        parton densities in the proton: GRV98\cite{epj:c5:461} 
\item        renormalization scale = factorization scale = $\sqrt{Q^2+4M_{c,b}^2}$ 
\item        $M_c=1.4 ; M_b=4.5\,\gev$
\end{itemize}

\begin{figure}[hbt]
  \begin{center}
    \epsfig{figure=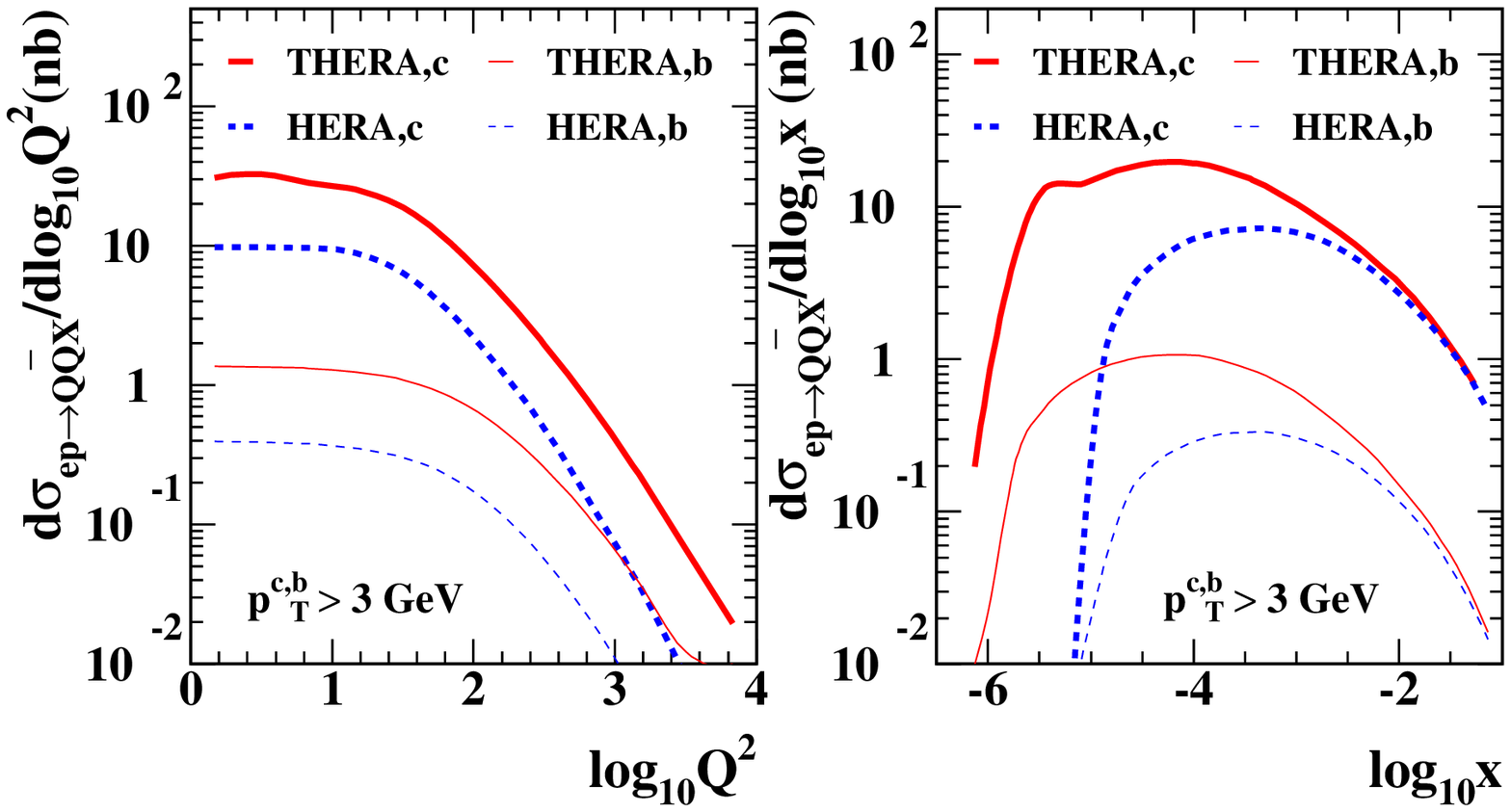,width=0.90\textwidth}
  \end{center}
  \vskip-5.2cm
  \centerline{\large\bf\kern3.5cm a)\kern8.cm b)\hfill}
  \vskip4.4cm
  \caption{
    The differential cross sections for charm (thick curves) and beauty (thin
    curves) production in neutral current DIS calculated in NLO QCD, (a)
    $d\sigma/d\log_{10}Q^2$ and (b) $d\sigma/d\log_{10}x$.
    The cross sections at
    THERA (solid red curves) and HERA (dashed blue curves) are compared.}
  \label{fig:hqpm:cball}
\end{figure}

\begin{figure}[hbt]
\vfill
  \begin{center}
    \centerline{\epsfysize 4 cm\mbox{\epsfig{figure=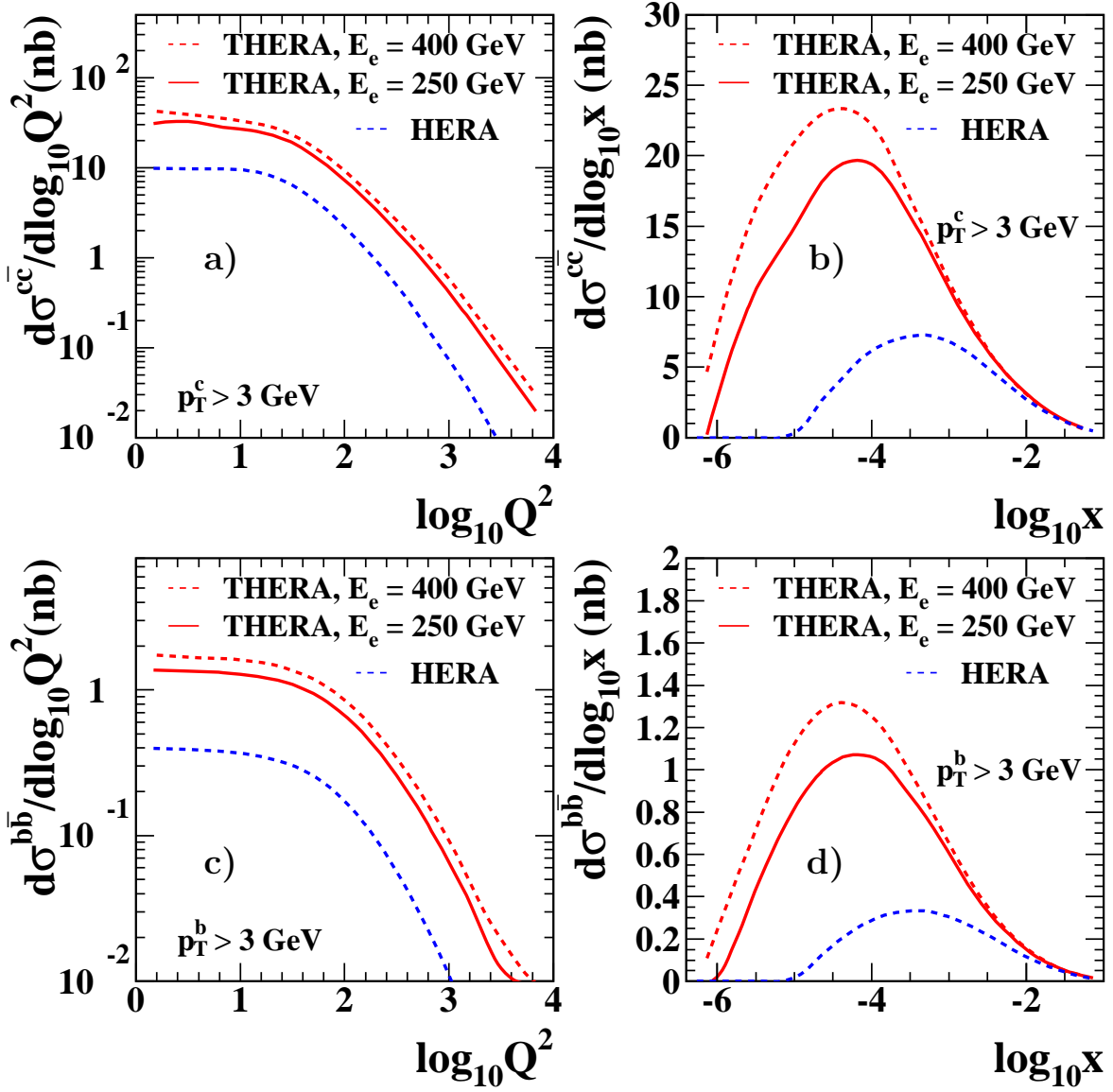,width=17cm}}}
  \end{center}
  \vskip-13.2cm
  \centerline{\large\bf\kern3.0cm a)\kern8.0cm b)\hfill}
  \vskip8.0cm
  \centerline{\large\bf\kern3.0cm c)\kern8.0cm d)\hfill}
  \vskip3.4cm
  \caption{The differential cross sections
   $d\sigma/d\log_{10}Q^2$ ((a) and (c)) and $d\sigma/d\log_{10}x$
   ((b) and (d))
   for charm ((a) and (b)) and beauty ((c) and (d))
   production at THERA (solid curves) and HERA (lower dashed curves).
   The upper dashed
   curves correspond to THERA operation with $E_e=400\,\gev$.}
  \label{fig:hqpm:cball400}
\vfill
\end{figure}

Fig.~\ref{fig:hqpm:cball} displays the differential
cross sections
$d\sigma/d\log_{10}Q^2$ and $d\sigma/d\log_{10}x$
for charm and beauty production at THERA and HERA
in the kinematic region defined above.
The most prominent difference is the increase of the cross sections
for both charm and beauty at low $x$. The cross sections
are experimentally measurable at values as low as $10^{-6}$. 
The effect of increasing the energy of the electron beam to $400\,\gev$
is shown in Fig.~\ref{fig:hqpm:cball400}.

\begin{figure}[hbt]
\vfill
  \begin{center}
    \centerline{\epsfysize 4 cm\mbox{\epsfig{figure=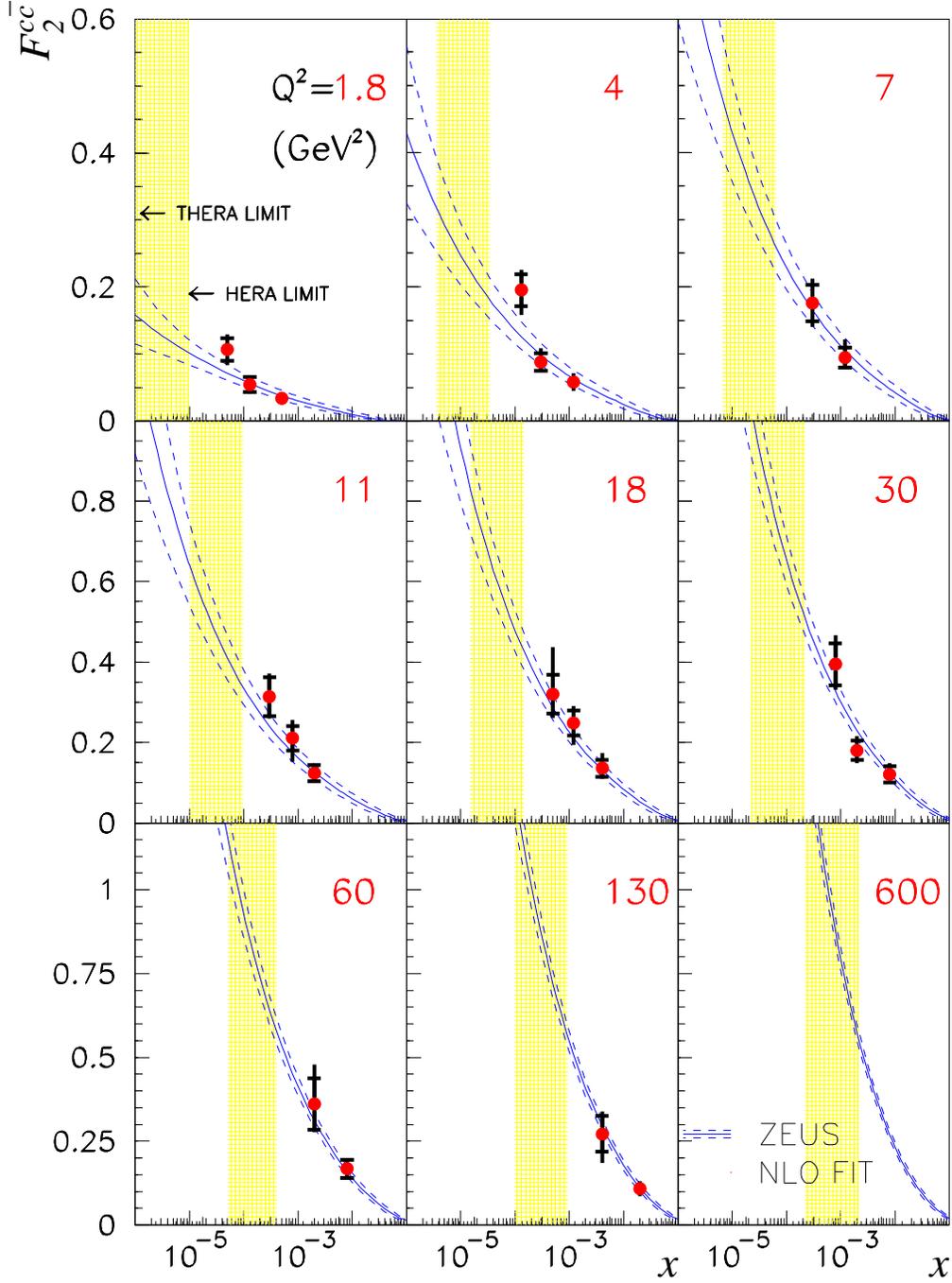}}}
  \end{center}
  \vskip-8.mm
  \caption{
  $F_2^{c\bar{c}}$ at $Q^2$ values between 1.8 and $600\,\gev^2$
  as a function of $x$.
  The curves correspond to the three flavour FFNS NLO QCD calculation
  using the parton distributions from the ZEUS NLO fit.
  The solid curves correspond to the central values and the dashed curves
  give the uncertainty due to the parton distributions from the ZEUS NLO fit.
  The  vertical shaded (yellow) bands show the small $x$ extension
  from HERA  to THERA for every $Q^2$.
  The $F_2^{c\bar{c}}$ values measured by the ZEUS collaboration
  are shown as an illustration.}
  \label{fig:hqpm:f2cc}
\vfill
\end{figure}

Fig.~\ref{fig:hqpm:f2cc} illustrates the extension of
the kinematic range of the $F_2^{c\bar{c}}$ measurement
from HERA to THERA.
$F_2^{c\bar{c}}$ at $Q^2$ values between 1.8 and $600\,\gev^2$
is plotted as a function of $x$.
The $F_2^{c\bar{c}}$ values measured
by the ZEUS collaboration~\cite{epj:c12:35} are shown as an illustration.
The expected extension is plotted as vertical shaded (yellow) bands.
The bands have been produced from the difference of the THERA
and HERA low $x$ limits obtained with HVQDIS.
The curves in Fig.~\ref{fig:hqpm:f2cc} correspond to 
the three flavour FFNS NLO QCD calculation~\cite{np:b392:162} using
the parton distributions from the ZEUS NLO QCD fit~\cite{epj:c7:609}.
The figure shows a gain of $\sim1$ order of magnitude in $x$
for all $Q^2$ values.
And, since $F_2^{c\bar{c}}$ and the gluon density  at larger $Q^2$ values
increase steeper towards small $x$, the gain in the accepted
$x$ range produces an increase of the ratio between THERA
and HERA cross sections with $Q^2$.
This ratio  rises from $\sim$3(3.5)  at low $Q^2$ to $\sim$5(5.5) at
$\log_{10}(Q^2)$=3.5  for charm (beauty) production.
Of course, to benefit from the rise and to improve substantially HERA results
in the high $Q^2$ region,
THERA luminosity should be of the same order
as the final HERA luminosity ($\sim1\,\text{fb}^{-1}$).

\begin{figure}[hbt]
\vfill
  \centerline{\mbox{\epsfig{figure=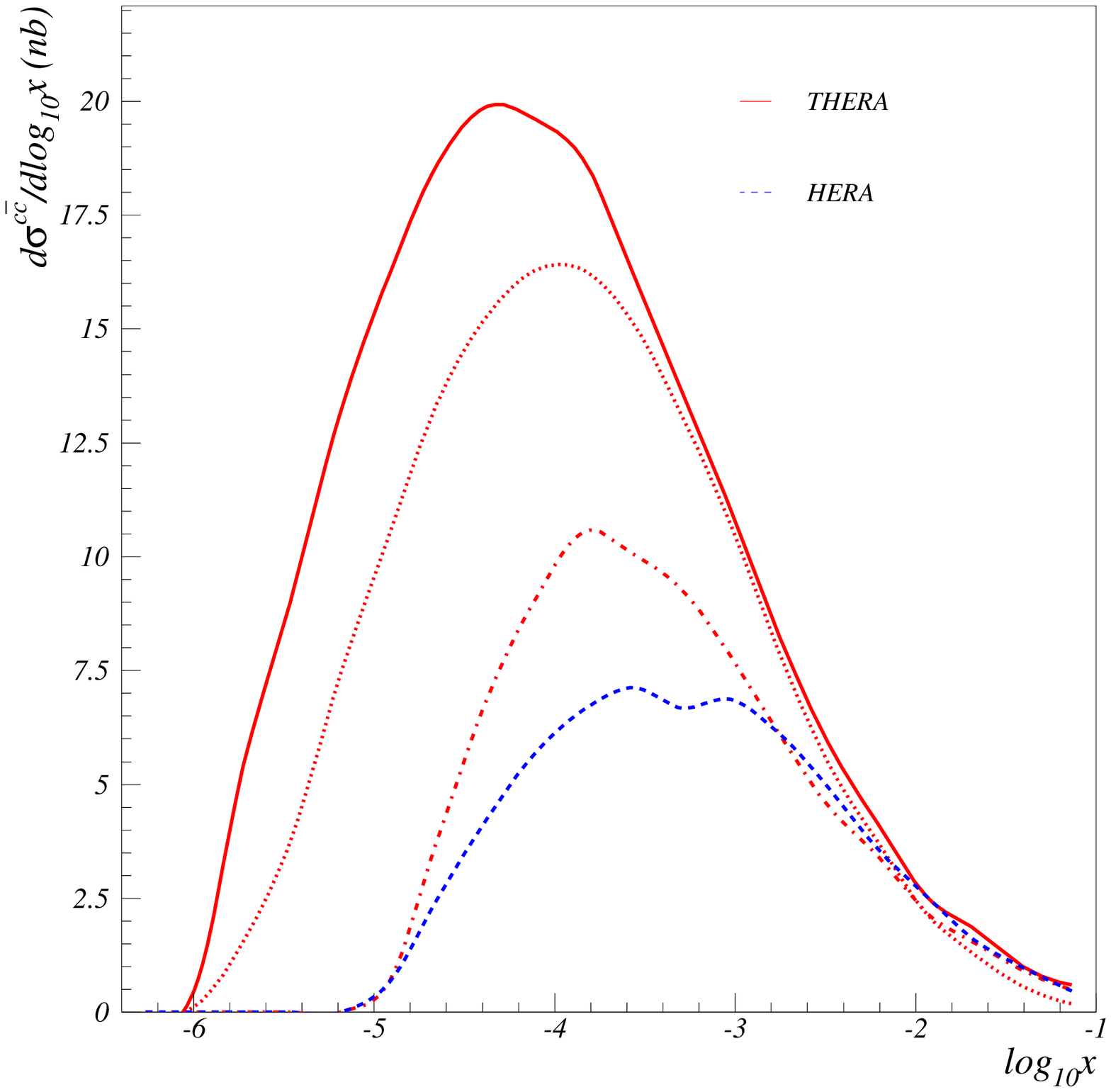,width=8cm}}
    \begin{minipage}[b]{8cm}
  \caption{
  The differential cross sections $d\sigma/d\log_{10}x$ for
  charm quark production at THERA (solid red curve) and
  HERA (dashed blue curve).
  The dash-dotted and dotted red curves show the cross sections for THERA
  with the additional cuts $Q^2>10\,\gev^2$ and  $\mid \eta^c \mid<$3,
  respectively.}
  \vspace{3cm}
  \end{minipage}}
\label{fig:hqpm:check}
\vfill
\end{figure}

The gain in the acceptance to low $x$ values will be marginal
if it will be impossible to measure DIS with $Q^2$ below $10\,\gev^2$
at THERA.
Fig.~\ref{fig:hqpm:check} shows the differential
cross sections for charm quark production at THERA (solid curve)
and HERA (dashed curve).
The effect of the $Q^2>10\,\gev^2$ requirement for the THERA case
is shown by the dash-dotted curve.
THERA sensitivity to low $x$ values depends also
on the $\eta$ range of the heavy quark tagging.
The dotted curve shows the THERA cross section with the cut
$\mid \eta^c \mid<$3 used instead of $\mid \eta^c \mid<$5.
Relaxing the minimum $p^{c}_\perp$ cut
leads to a gain in the accepted THERA cross section but not in
the low $x$ reach.

\clearpage

\section{Charm production in charged current DIS}

\begin{figure}[hbt]
  \begin{center}
    \epsfig{figure=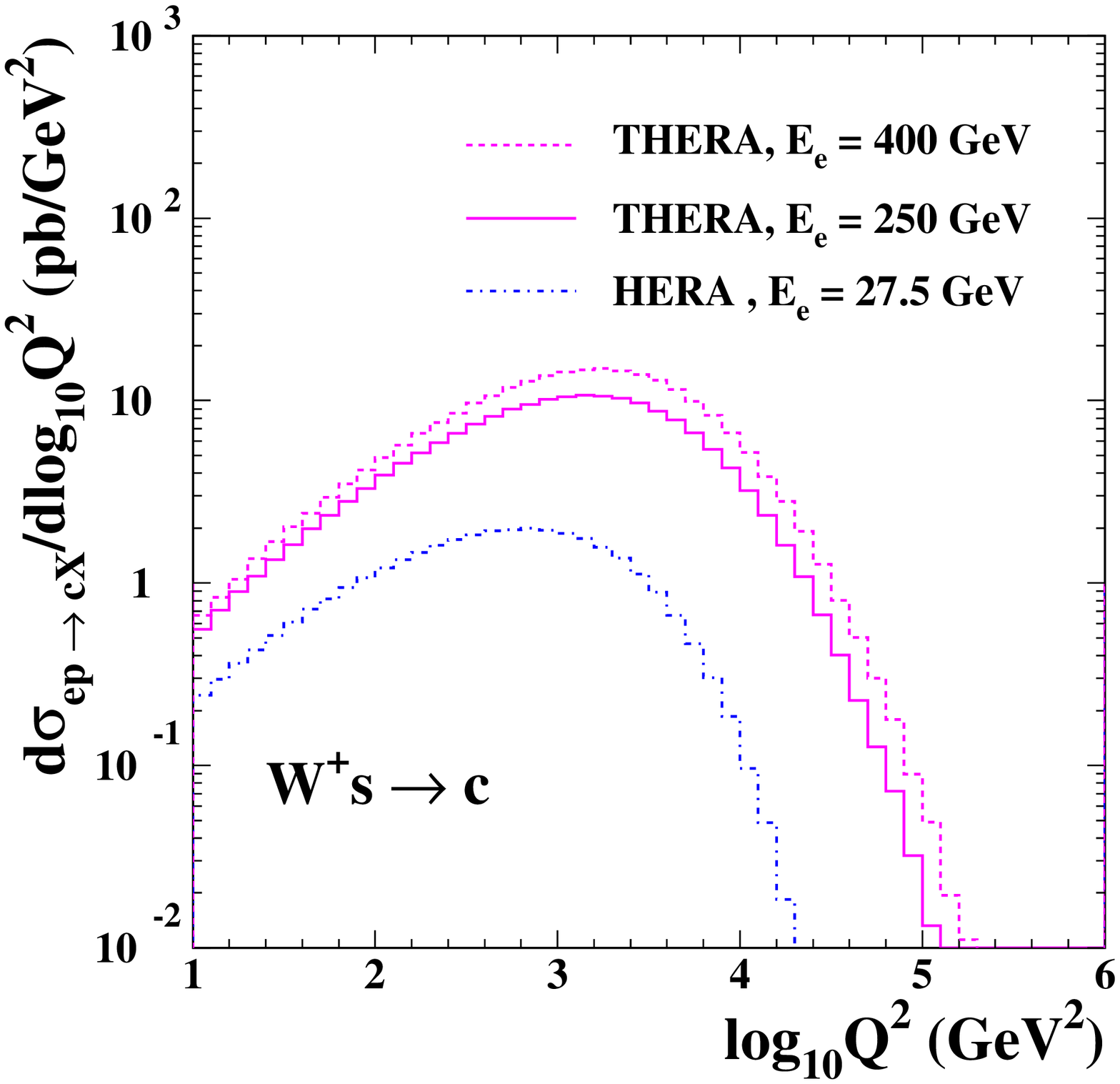,width=7.5cm}
    \epsfig{figure=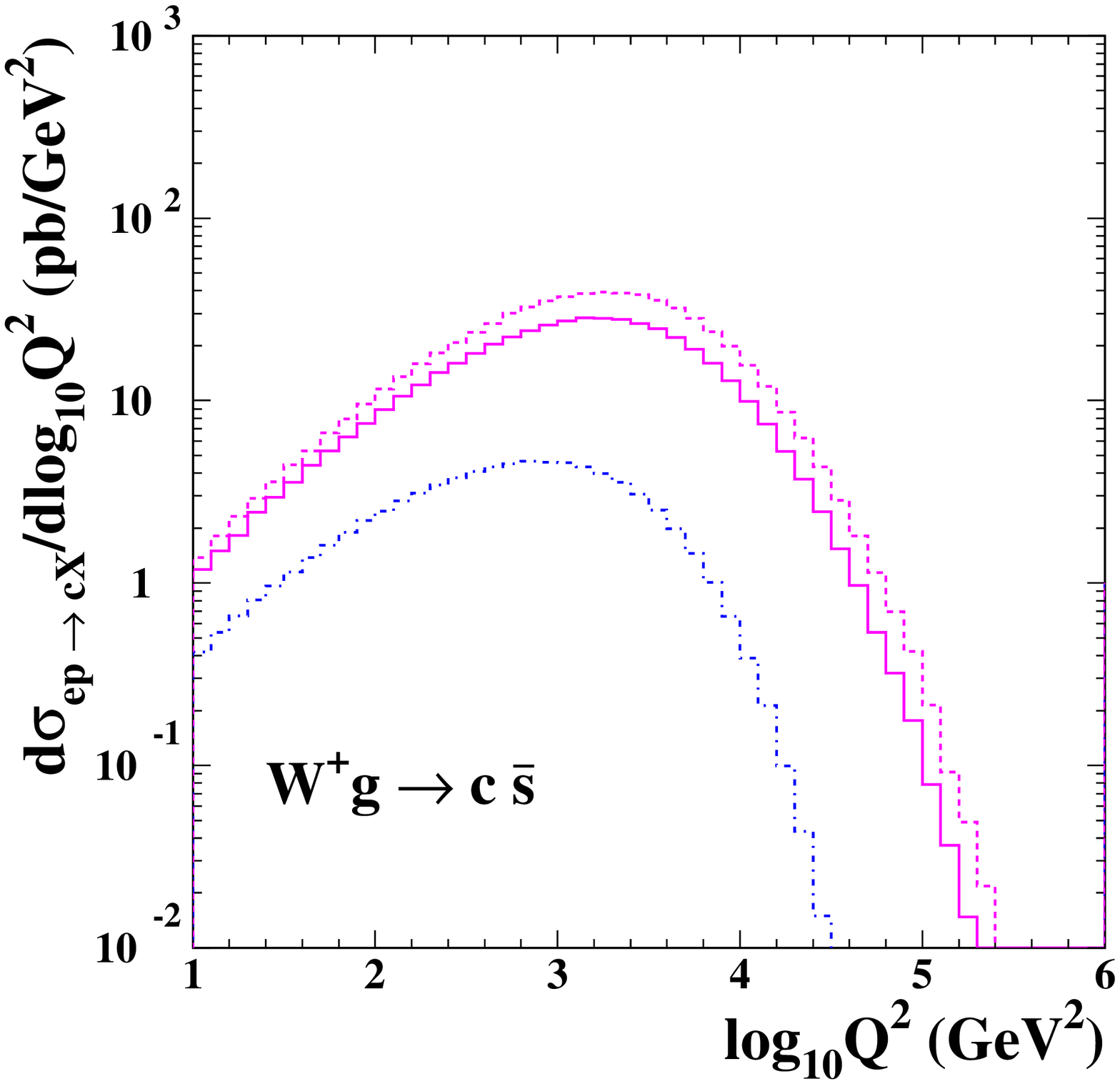,width=7.5cm}
  \end{center}
  \vskip-6.5cm
  \centerline{\large\bf\kern2.cm a)\kern7.5cm b)\hfill}
  \vskip5.7cm
\begin{center}
\epsfig{figure=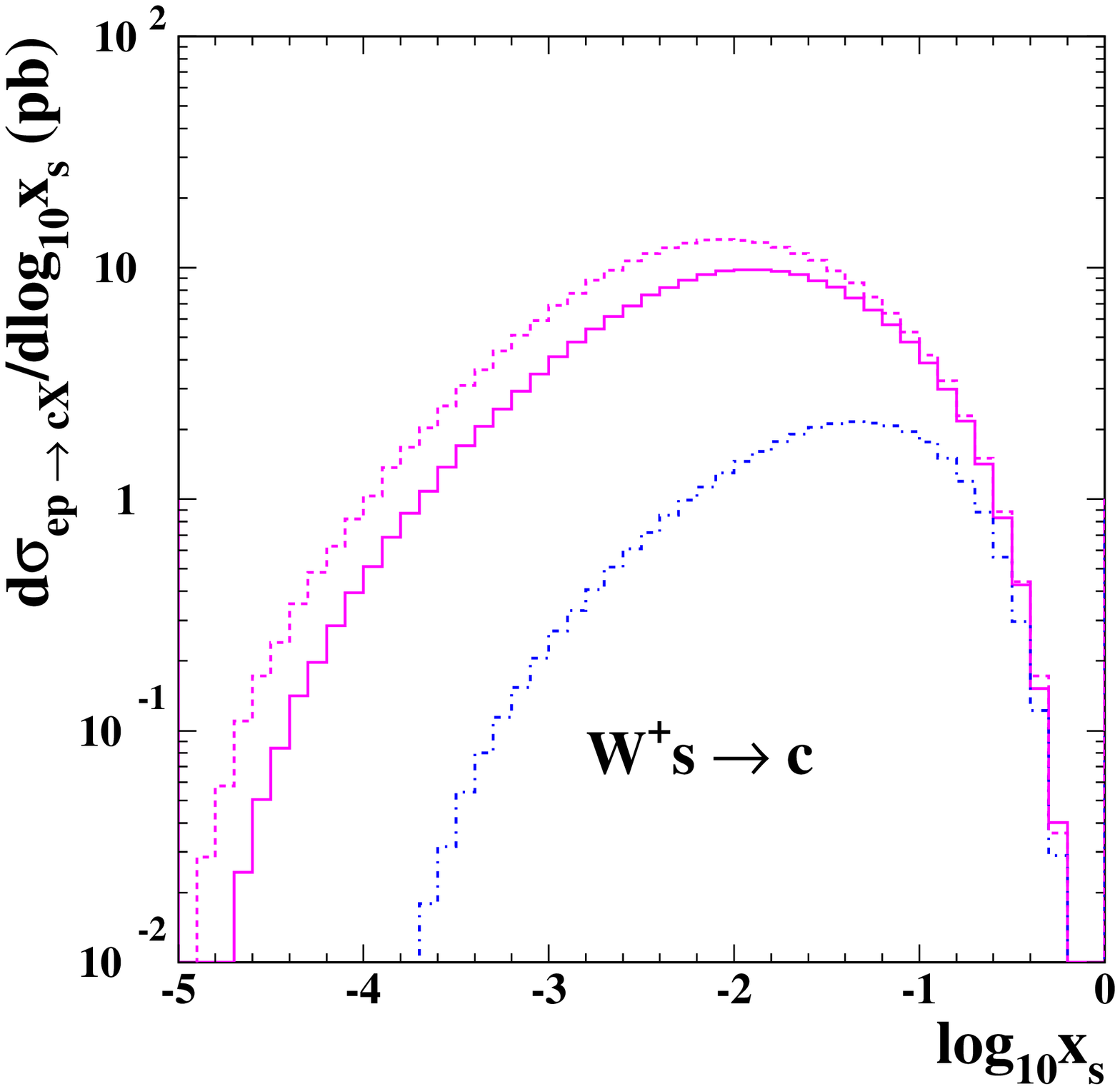,width=7.5cm}
\epsfig{figure=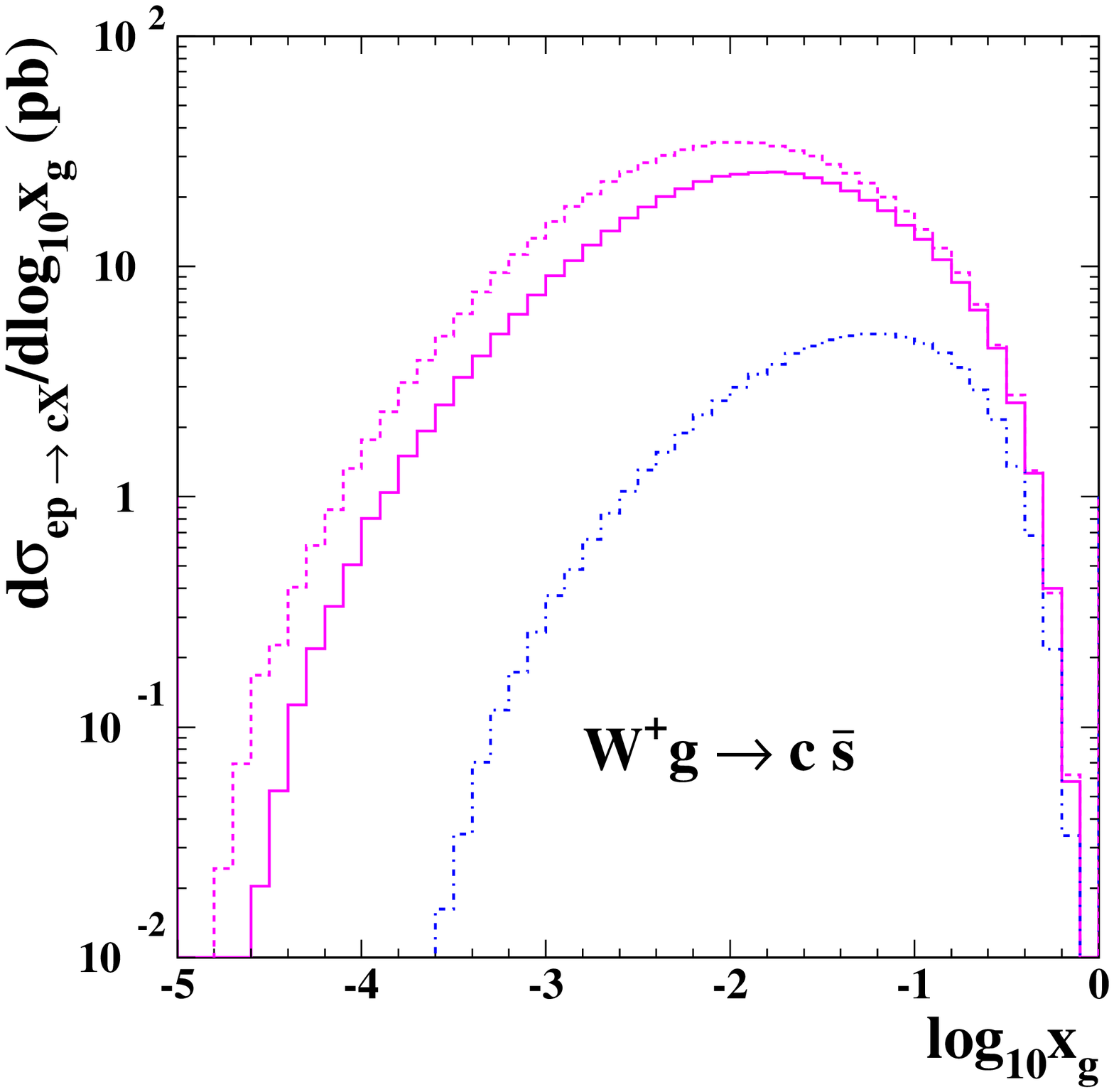,width=7.5cm}
\end{center}
  \vskip-6.5cm
  \centerline{\large\bf\kern2.cm c)\kern7.5cm d)\hfill}
  \vskip5.7cm
  \caption{
    The differential cross sections $d\sigma/d\log_{10}Q^2$ and
    $d\sigma/d\log_{10}x_{s,g}$
    for charm production in charged current DIS
    from the strange sea ((a) and (c)) and from boson--gluon
    fusion ((b) and (d)),
    calculated with the LO Monte Carlo generator HERWIG.
    The solid and dashed magenta curves show the
    predictions for THERA operation with an electron energy of $250\,\gev$ and
    $400\,\gev$, respectively, and $E_p=920\,\gev$. The predictions
    for the HERA
    case are indicated by the dash-dotted blue curves.}
  \label{fig:hqpm:cchf}
\end{figure}

\begin{figure}[hbt]
  \begin{center}
    \epsfig{figure=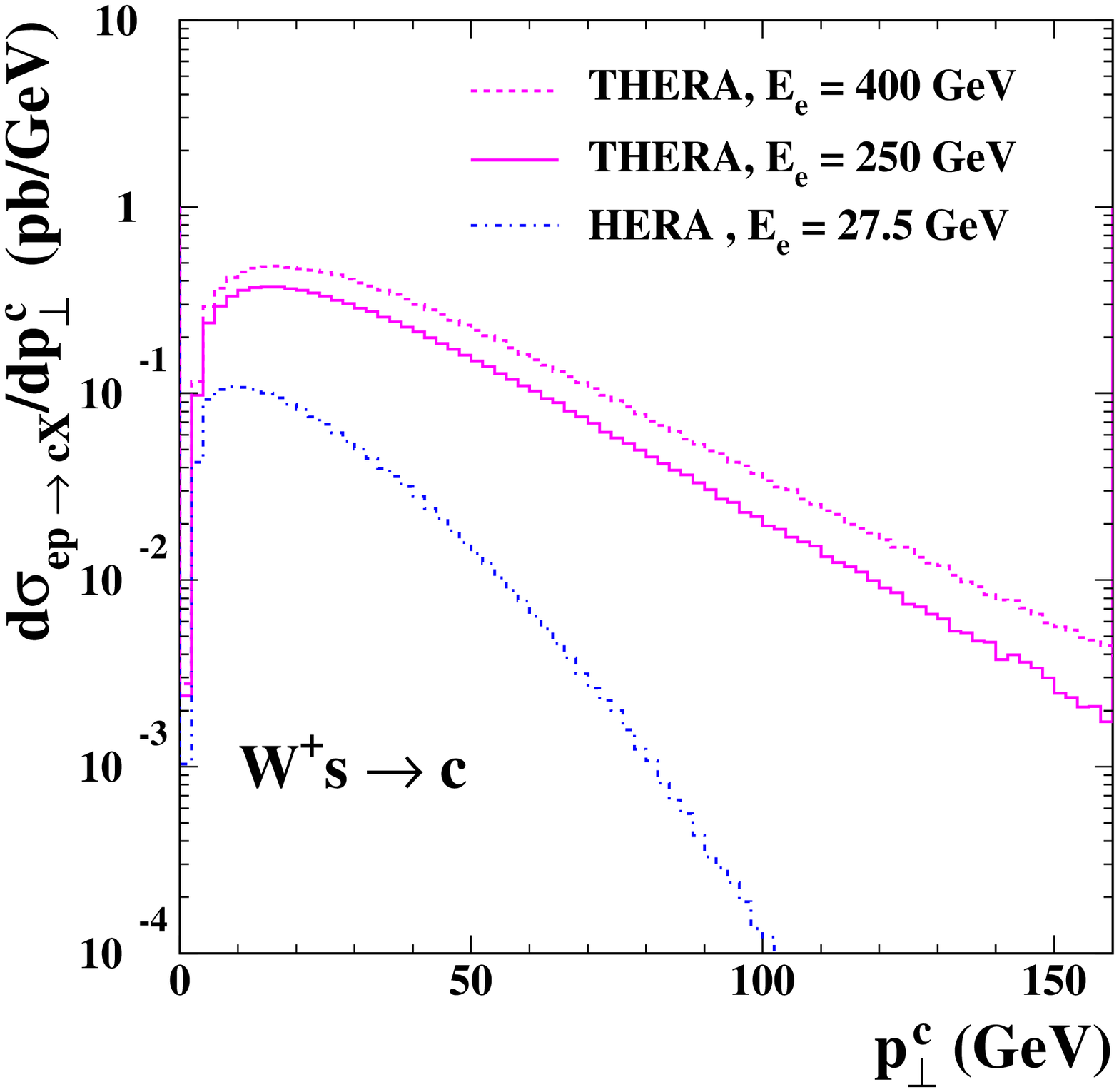,width=7.5cm}
    \epsfig{figure=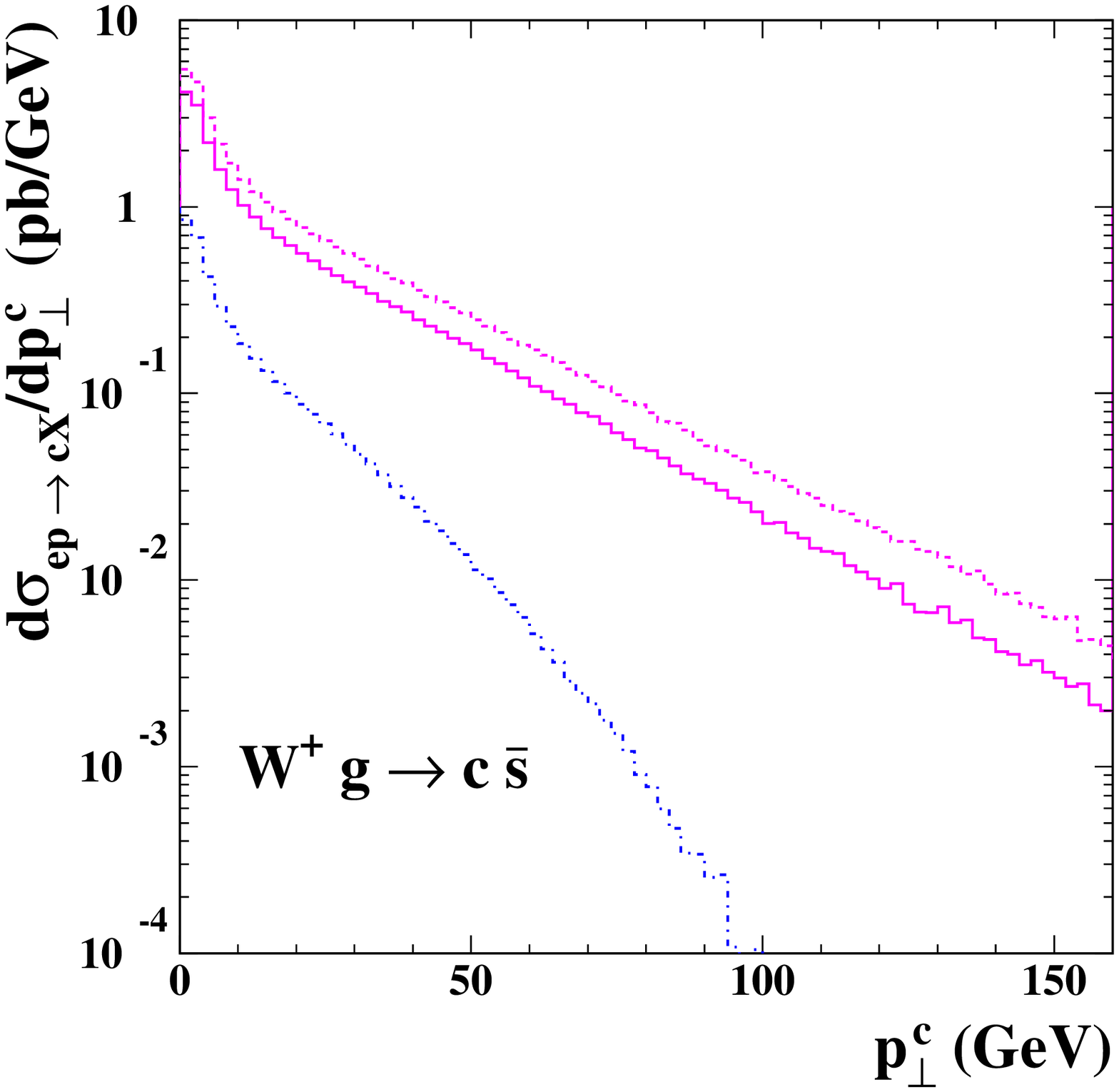,width=7.5cm}
  \end{center}
  \vskip-6.7cm
  \centerline{\large\bf\kern2.cm a)\kern7.5cm b)\hfill}
  \vskip5.7cm
\begin{center}
\epsfig{figure=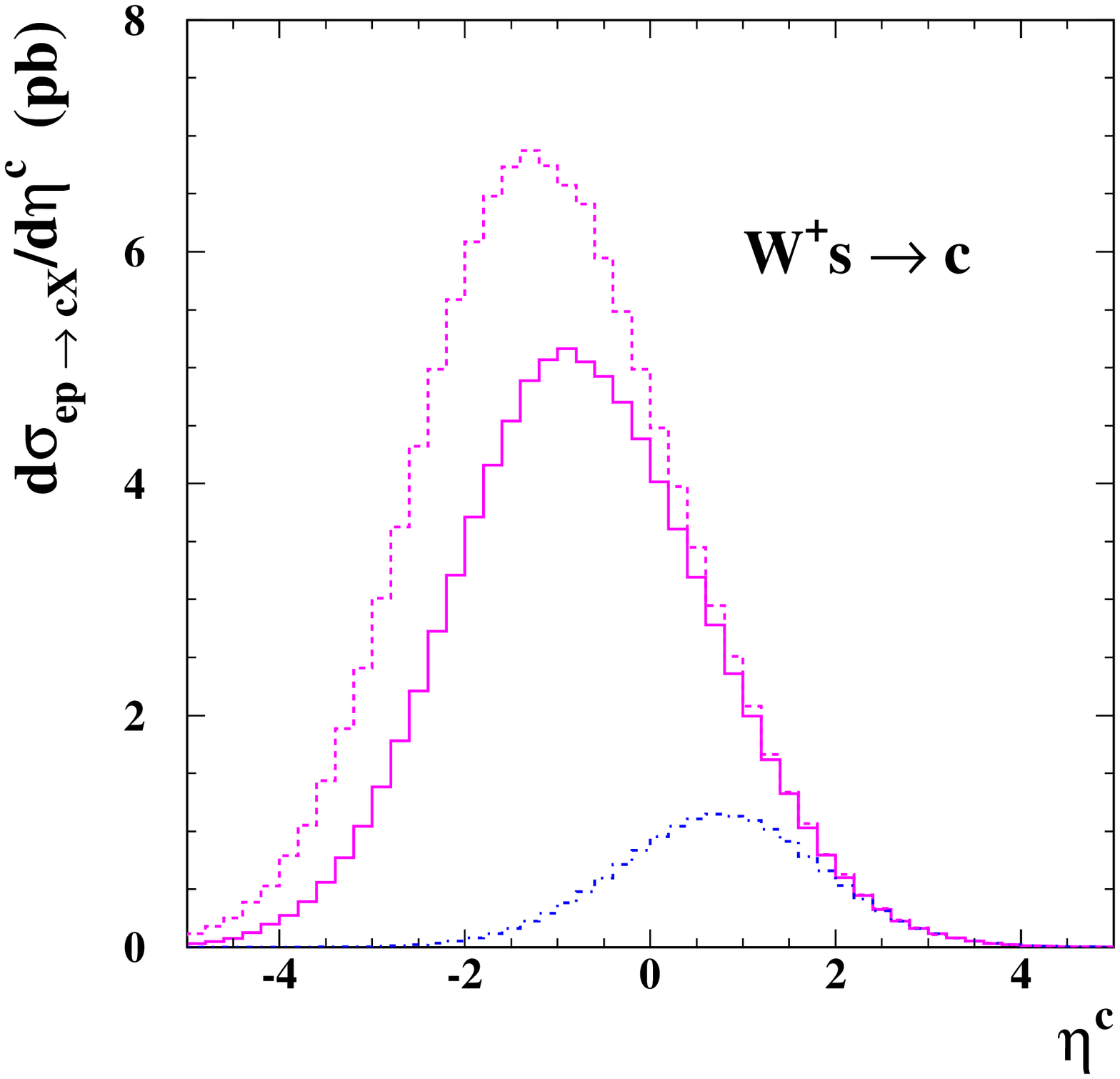,width=7.5cm}
\epsfig{figure=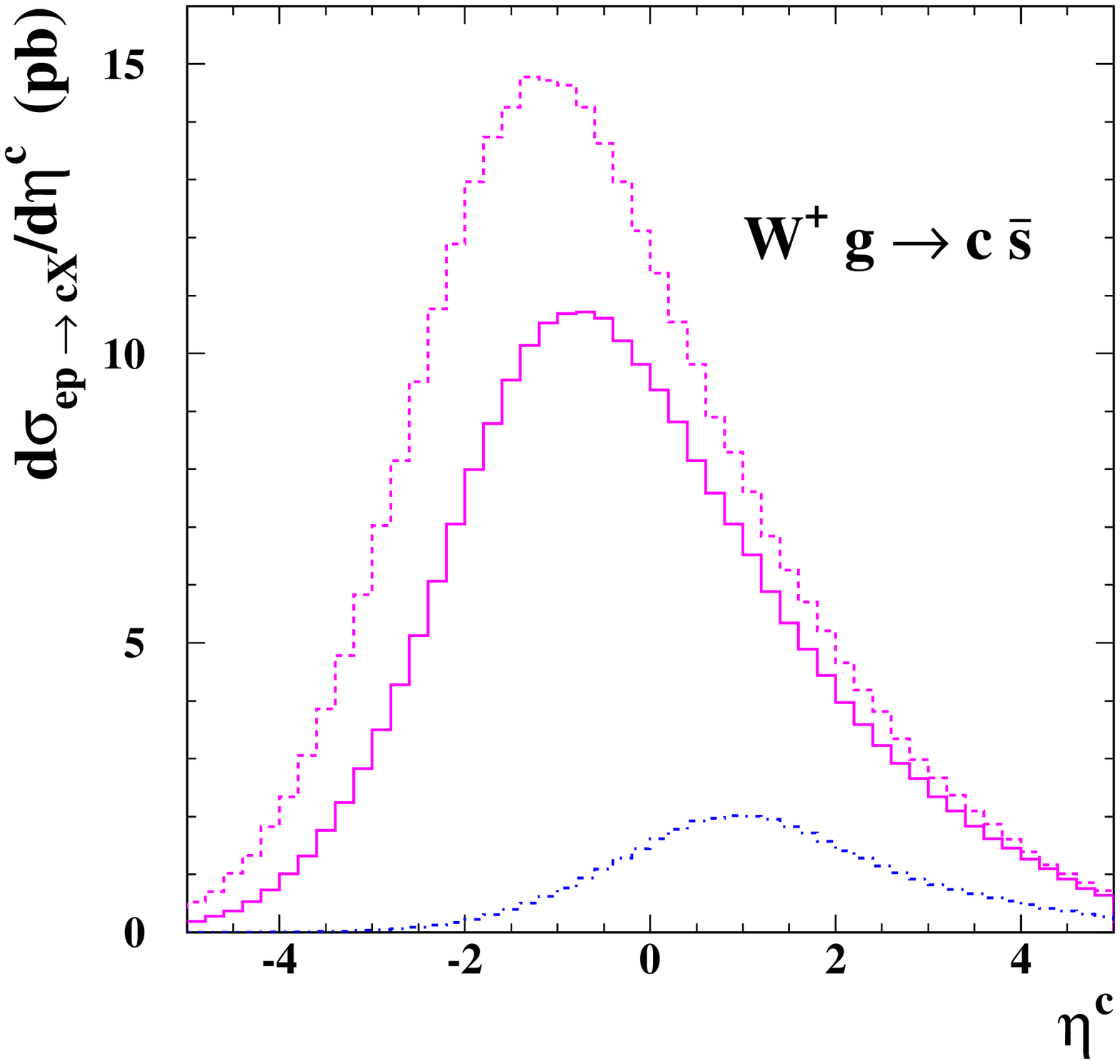,width=7.5cm}
\end{center}
  \vskip-6.6cm
  \centerline{\large\bf\kern2.cm c)\kern7.5cm d)\hfill}
  \vskip5.7cm
  \caption{
    The differential cross sections $d\sigma/dp^c_\perp$ and
    $d\sigma/d\eta^c$
    for charm production in charged current DIS from
    the strange sea ((a) and (c)) and from boson--gluon
    fusion ((b) and (d)), calculated with the LO Monte Carlo
    generator HERWIG.    
    The solid and dashed magenta curves show the
    predictions for THERA operation with an electron energy of $250\,\gev$ and
    $400\,\gev$, respectively, and $E_p=920\,\gev$. The predictions
    for the HERA
    case are indicated by the dash-dotted blue curves.}
  \label{fig:hqpm:cchf_pt}
\end{figure}

The theoretical description of charm production in charged current (CC) DIS is
challenging~\cite{np:b500:301,*hep-ph-0010344}.  The special interest in this
process is caused by its sensitivity to the proton strange-quark density which
is rather poorly known~\cite{hep-ph-0004268}.  However, no measurements of CC
charm production have been performed at HERA so far due to the small
process cross section ($\sim10\,\text{pb}$).
According to the LO Monte Carlo HERWIG calculation, the cross
sections for both LO CC charm production processes, $W^+s\to c$ and $W^+g\to
c\bar{s}$, will be more than 6~times larger at THERA than at HERA.
The differential cross sections $d\sigma/d\log_{10}Q^2$ and
$d\sigma/d\log_{10}x_{s,g}$
($x_{s,g}$ denoting the parton fractional momenta in the proton)
for CC charm production are shown in Fig.~\ref{fig:hqpm:cchf}.
The THERA cross sections are shifted towards
larger $Q^2$ with respect to those at HERA.  They are one order of magnitude
larger than the HERA cross sections at large $Q^2$ values, thereby creating the
opportunity to study charm production in CC DIS at THERA.
Fig.~\ref{fig:hqpm:cchf} shows also that
charm production in CC at THERA is sensitive to much wider
ranges in $x_{s,g}$
with respect to those at HERA.

The separation of the strange see contribution to
the charm production in CC will require a special experimental
procedure. It could utilize the different
kinematics of $W^+s\to c$ and $W^+g\to c\bar{s}$ processes.
Fig.~\ref{fig:hqpm:cchf_pt} shows the differential cross sections
$d\sigma/ dp^{c}_\perp$ and $d\sigma/ d\eta^{c}$ for the processes
at HERA and THERA.
The pseudorapidity distributions are rather close for both
processes while the $p^{c}_\perp$ distributions are
remarkably different. The boson--gluon fusion component is a few orders
of magnitude
larger than the strange see component at small $p^{c}_\perp$ values.
At large $p^{c}_\perp$ values both component contributions are similar. 
Tagging charm quarks with $p^{c}_\perp$
above a few $\gev$
suppresses effectively
the boson--gluon fusion component contribution.
Further separation of the strange see contribution will probably require
a selection of events with only one jet representing the charm quark.

\section{Summary}

The total cross sections of charm and beauty production at THERA
are expected to increase by factors of three and five, respectively,
as compared to HERA.
Heavy quarks can be measured at THERA in wide ranges of their
transverse momenta,
thereby providing a solid basis for testing
the pQCD calculations.
The gluon structure of the proton can be probed at THERA in
as yet unexplored ranges.
The kinematic limits of the $x_g$ measurements are
$10^{-5}$ and $10^{-4}$ for charm and beauty production, respectively.
The measurements will require
the special tracking and muon identification devices
in the backward (electron) direction.

THERA will open new regions for the heavy quark DIS production
never measured before.
For all $Q^2$ values the kinematic limit in $x$ at THERA
is $\sim1$ order of magnitude smaller
with respect to that at HERA.
Measuring the extreme low $x$ ($10^{-5}$--$10^{-6}$)
regime is an experimental challenge.
If the luminosity turns out to be comparable with final HERA numbers,
beauty and charm production at high $Q^2$ values will benefit from
the $3-5$ times larger cross sections at THERA.

The cross section of charm production in charged current at THERA
is $\sim1$ order of magnitude larger than that at HERA,
thereby creating the opportunity to study the process.

\clearpage

\section*{Acknowledgments}

We would like to thank S.~Frixione and B.~Harris for providing us
with the programs for their NLO calculations.

\def\bibname{References}
\def\refname{References}
\bibliographystyle{./thera}
{\raggedright
\bibliography{./thera,./gladilin}

\providecommand{\etal}{et al.\xspace}
\providecommand{\coll}{Coll.}
\catcode`\@=11
\def\@bibitem#1{%
\ifmc@bstsupport
  \mc@iftail{#1}%
    {;\newline\ignorespaces}%
    {\ifmc@first\else.\fi\orig@bibitem{#1}}
  \mc@firstfalse
\else
  \mc@iftail{#1}%
    {\ignorespaces}%
    {\orig@bibitem{#1}}%
\fi}%
\catcode`\@=12
\begin{mcbibliography}{10}

\bibitem{pl:b308:137}
S.~Frixione et al.,
\newblock Phys.~Lett.{} {\bf B~308},~137~(1993)\relax
\relax
\bibitem{np:b545:21}
H1 Coll., C.~Adloff et al.,
\newblock Nucl.~Phys.{} {\bf B~545},~21~(1999)\relax
\relax
\bibitem{epj:c6:67}
ZEUS Coll., J.~Breitweg et al.,
\newblock Eur.~Phys.~J.{} {\bf C~6},~67~(1999)\relax
\relax
\bibitem{epj:c12:35}
ZEUS Coll, J.~Breitweg et al.,
\newblock Eur.~Phys.~J.{} {\bf C~12},~35~(2000)\relax
\relax
\bibitem{pl:b481:213}
ZEUS Coll., J.~Breitweg et al.,
\newblock Phys.~Lett.{} {\bf B~481},~213~(2000)\relax
\relax
\bibitem{pl:b348:633}
S.~Frixione et al.,
\newblock Phys.~Lett.{} {\bf B~348},~633~(1995)\relax
\relax
\bibitem{np:b454:3}
S.~Frixione et al.,
\newblock Nucl.~Phys.{} {\bf B~454},~3~(1995)\relax
\relax
\bibitem{zfp:c76:689}
B.~A.~Kniehl et al.,
\newblock Z.~Phys.{} {\bf C~76},~689~(1997)\relax
\relax
\bibitem{pr:d58:014014}
J.~Binnewies et al.,
\newblock Phys.~Rev.{} {\bf D~58},~014014~(1998)\relax
\relax
\bibitem{pr:d55:2736}
M.~Cacciari et al.,
\newblock Phys.~Rev.{} {\bf D~55},~2736~(1997)\relax
\relax
\bibitem{pr:d55:7134}
M.~Cacciari et al.,
\newblock Phys.~Rev.{} {\bf D~55},~2736~(1997)\relax
\relax
\bibitem{pr:d57:2806}
B.W.~Harris and J.~Smith,
\newblock Phys.~Rev.{} {\bf D~57},~2806~(1998)\relax
\relax
\bibitem{pr:d50:3102}
M. A. G. Aivazis {\it et al.},
\newblock Phys.\ Rev.{} {\bf D~50},~3102~(1994)\relax
\relax
\bibitem{jhep:0010:031}
J.~Amundson et al.,
\newblock JHEP{} {\bf 0010},~031~(2000).
\newblock Also in hep-ph/0005221\relax
\relax
\bibitem{pr:d61:096004}
A.~Chuvakin, J.~Smith, W.~L.~van~Neerven,
\newblock Phys.\ Rev.{} {\bf D~61},~096004~(2000)\relax
\relax
\bibitem{pl:b467:156}
H1 Coll., C.~Adloff et al.,
\newblock Phys.~Lett.{} {\bf B~467},~156~(1999)\relax
\relax
\bibitem{epj:c18:625}
ZEUS Coll., J.~Breitweg et al.,
\newblock Eur.~Phys.~J.{} {\bf C~18},~625~(2001)\relax
\relax
\bibitem{np:b412:225}
S.~Frixione et al.,
\newblock Nucl.~Phys.{} {\bf B~412},~225~(1994)\relax
\relax
\bibitem{np:b373:295}
M.L.~Mangano, P.~Nason and G.~Ridolfi,
\newblock Nucl.~Phys.{} {\bf B~373},~295~(1992)\relax
\relax
\bibitem{h1-bdis}
H1 Coll.,
\newblock {\em Beauty production in deep inelastic scattering}.
\newblock Presented by T.~Sloan at XXXVIth Rencontres de Moriond: QCD, Les
  Arcs, France, March 2001\relax
\relax
\bibitem{cpc:67:465}
G.~Marchesini et al.,
\newblock Comp.~Phys.~Comm.{} {\bf 67},~465~(1992)\relax
\relax
\bibitem{thera-jankowski}
P.~Jankowski, M.~Krawczyk and M.~Wing,
\newblock {\em Heavy quark photoproduction at {THERA}},
\newblock in {\em The THERA Book, DESY-LC-REV-2001-062 (2001)}.
\newblock Also in hep-ph/0103330\relax
\relax
\bibitem{thera-baranov}
S.P.~Baranov and N.P.~Zotov,
\newblock {\em Heavy quark production in the semihard approach at {THERA}},
\newblock in {\em The THERA Book, DESY-LC-REV-2001-062 (2001)}.
\newblock Also in hep-ph/0103138\relax
\relax
\bibitem{np:b392:162}
E.~Laenen, S.~Riemersma, J.~Smith and W.~L.~van~Neerven,
\newblock Nucl.\ Phys.{} {\bf B~392},~162~(1995)\relax
\relax
\bibitem{pl:b353:535}
B.~W.~Harris and J.~Smith,
\newblock Phys.\ Lett.{} {\bf B~353},~535~(1995)\relax
\relax
\bibitem{epj:c5:461}
M.~Gl\"uck, E.~Reya and A.~Vogt,
\newblock Eur.~Phys.~J.{} {\bf C~5},~461~(1998)\relax
\relax
\bibitem{epj:c7:609}
ZEUS Coll., J.~Breitweg et al.,
\newblock Eur.~Phys.~J.{} {\bf C~7},~609~(1999)\relax
\relax
\bibitem{np:b500:301}
M.~Buza and W.L.~van~Neerven,
\newblock Nucl.~Phys.{} {\bf B~500},~301~(1997)\relax
\relax
\bibitem{hep-ph-0010344}
R.S.~Thorne and R.G.~Roberts,
\newblock {\em A variable number flavor scheme for charged current heavy flavor
  structure functions}.
\newblock Preprint \mbox{RAL-TR-2000-048} (\mbox{hep-ph/0010344}), 2000\relax
\relax
\bibitem{hep-ph-0004268}
V.~Barone, C.~Pascaud and F.~Zomer,
\newblock {\em A new global analysis of {DIS} data and the strange sea
  distribution},
\newblock in {\em Proc.\ Workshop on Light-Cone QCD and Nonperturbative Hadron
  Physics, Adelaide, Australia, 1999}, p. 167.
\newblock 2000.
\newblock Also in preprint \mbox{hep-ph/0004268}\relax
\relax
\end{mcbibliography}
}

\end{document}